\documentstyle[psfig,amssymb]{mn2e}
\begin{document}

\newif\ifAMStwofonts
\topmargin -0.5cm 

\title[Mass-to-light ratios in fossil groups]
{On the mass-to-light ratios of fossil groups. Are they simply dark clusters?}
\author[
Proctor \rm{et al.} 
]
{Robert N. Proctor$^{1}$, Claudia Mendes de Oliveira$^{1}$, Renato Dupke$^{2,3,4}$, \\
\\
\LARGE  Raimundo Lopes de Oliveira$^{5,8}$, Eduardo S. Cypriano$^{1}$, Eric D. Miller$^6$, Eli Rykoff$^7$\\
\\
$^1$  Universidade de S\~{a}o Paulo, IAG, Rua do Mat\~{a}o, 1226, S\~{a}o Paulo, 05508-900, Brasil\\
$^2$ University of Michigan, Ann Arbor, MI, 48109, USA\\
$^3$ Eureka Scientific Inc., Oakland, CA 94602-3017\\
$^4$ Observat\'{o}rio Nacional, Rua Gal. Jos\'{e} Cristino, 20921-400, Rio de Janeiro, Brazil\\
$^5$ Universidade de S\~{a}o Paulo, Instituto de F\'{i}sica de S\~{a}o Carlos,
Caixa Postal 369, 13560-970, S\~{a}o Carlos, SP, Brazil \\
$^6$ Kavli Institute for Astrophysics and Space Research, Massachusetts Institute of Technology, Cambridge, MA 02139 \\
$^7$ E. O. Lawrence Berkeley National Lab, 1 Cyclotron Rd. Berkeley CA, 94720\\
$^8$ Universidade Federal de Sergipe, Departamento de F\'{i}sica, Av. Marechal Rondon s/n, 49100-000 S\~{a}o Crist\'{o}v\~{a}o, SE, Brazil\\
Email: rproctor@astro.iag.usp.br}
\def\LaTeX{L\kern-.36em\raise.3ex\hbox{a}\kern-.15em
    T\kern-.1667em\lower.7ex\hbox{E}\kern-.125emX}

\label{firstpage}

\newpage

\maketitle

\begin{abstract}
Defined as X-ray bright galaxy groups with large differences between
the luminosities of their brightest and second brightest galaxies,
``fossil groups'' are believed to be some of the oldest galaxy systems
in the universe. They have therefore been the subject of much recent
research.

In this work we present a study of 10 fossil group candidates with an
average of 33 spectroscopically confirmed members per group, making
this the deepest study of its type to-date. We also use this data to
perform an analysis of the luminosity function of our sample of fossil
groups.

We confirm the high masses previously reported for many of fossil
systems, finding values more similar to those of clusters than of
groups. We also confirm the high dynamical mass-to-light ratios
reported in many previous studies. While our results are consistent
with previous studies in many ways, our interpretation is not. This is
because we show that, while the luminosities of the BCGs in these
systems are consistent with their high dynamical masses, their
richnesses (total number of galaxies above some canonical value) are
extremely low. This leads us to suggest a new interpretation of fossil
systems in which the large differences between the luminosities of
their brightest and second brightest galaxies are simply the result
the high BCG luminosities and low richnesses, while the high masses
and low richnesses also explain the high mass-to-light ratios. Our
results therefore suggest that fossil systems can be characterised as
cluster-like in their masses and BCG luminosities, but possessing the
richnesses and optical luminosities of relatively poor groups.
These findings are not predicted by any of the current models for the
formation of fossil groups. Therefore, if this picture is confirmed,
current ideas about the formation and evolution of fossil systems will
need to be reformulated.

\end{abstract}


\section{Introduction}
The study of galaxy groups and clusters has become a powerful tool in
many aspects of astrophysical research. From the cosmological
perspective, groups and clusters mark the most over-dense regions of
the matter distribution. They can therefore be used to constrain
cosmological parameters such as $\Omega_m$, $\sigma_8$ and {\it w} (the
equation of state of dark energy).

From the galaxy formation and evolution perspective, the low velocity
dispersions in galaxy groups result in frequent strong interactions
between galaxies (i.e. tidal disruption and merging).  The high
velocity dispersions in clusters, on the other hand, suppress strong
interactions between galaxies.  However, the deeper potential wells
and higher velocities in clusters mean that interactions with the
ambient environment (the cluster potential and the intra-cluster
medium) increase in importance, giving rise to processes such as
ram-pressure stripping and strangulation (e.g. Gunn \& Gott 1972;
Fujita 2004; Rasmussen, Ponman \& Mulchaey 2006; Kawata \& Mulchaey
2008).

Groups and clusters therefore provide an important testing ground for
models of galaxy formation and evolution, as well as enabling the
constraint of cosmological parameters. Consequently, there is an
ongoing effort to identify and characterise clusters and groups in
both the local universe and, more recently, at higher redshifts (Bauer
\rm{et al.} 2011; Hilton \rm{et al.} 2010; Strazzullo 2010).

A special class of groups/clusters, first identified by Ponman et
al. (1994), are ``fossil groups''. These are defined as X-ray luminous
structures ($L_X > 5 \times 10^{41} h_{70}^{-2}$ ergs s$^{-1}$) with a
greater than 2 magnitude gap between the brightest and second
brightest galaxies within half the virial radius (Jones et
al. 2003). Fossil groups are therefore dominated by a massive central
early-type galaxy surrounded by a swarm of much smaller galaxies and
enclosed in a hot X-ray halo.

The most commonly quoted scenario for the formation of such systems is
that, as a result of having remained relatively undisturbed for a
significant fraction of a Hubble time, dynamical friction has had time
to cause any large galaxies close to the central regions of the group
to spiral inwards, ultimately to merge with the central galaxy
(D'Onghia et al. 2005; Dariush \rm{et al.} 2007). This process
simultaneously increases the luminosity of the central galaxy and
depletes the central regions of massive (bright) galaxies, thus
creating the large luminosity gap which, by definition, characterises
FGs. However, there is as yet no direct evidence for this scenario.
Consequently, there is to date no consensus on the formation mechanism.

In an effort to address this issue, a few previous studies have
investigated the dynamical, X-ray and optical scaling relations of
fossil groups (e.g. Vikhlinin \rm{et al.}  1999; Jones \rm{et al.}
2003; Yoshioka \rm{et al.}  2004; Khosroshahi, Ponman \& Jones 2007,
hereafter KPJ07; Voevodkin \rm{et al.} 2010). Many find FGs to be more
X-ray luminous than non-fossil groups of the same optical luminosity,
while still following the same L$_X$-T$_X$ relation. This results in
most of these studies finding fossils to have very high mass-to-light
ratios. Each of these studies is at pains to point out that this is
consistent with their early formation - regardless of their preferred
formation mechanism. There are, however, two works in the literature
(Voevodkin \rm{et al.} 2010; Aguerri \rm{et al.} 2011) that refute
this X-ray excess. Voevodkin et al (2010) claim instead that ``the
X-ray brightness of massive fossil systems is consistent with that of
the general population of galaxy clusters and follows the same
L$_X$-–L$_{opt}$ scaling relation''. However, in a recent
re-examination of their data the authors found a serious error in
their estimations of the optical luminosities (private communication).
As a result, the authors agree that this finding is rendered
invalid. In the case of Aguerri \rm{et al.} (2011) the single system
reported is at a redshift of 0.5 and possesses a mass significantly
larger than any previously reported fossil system. In addition,
Aguerri \rm{et al.} report only a lower limit on the mass-to-light
ratio of the system. These issues make a meaningful comparison with
the low redshift systems in the literature problematic. The general
consensus therefore remains that fossil systems exhibit high
mass-to-light ratios.


In this paper we present an analysis of the dynamical, X-ray and
optical properties of ten fossil groups (five new and five previously
reported in the literature). Our study is similar to that of
KPJ07. However, due to our deeper spectroscopy, our sample comprises
an average of 33 galaxies per group, compared to $\sim$10 in
KPJ07. The large number of galaxies per group in our study also allows
us to look for spatio-dynamic substructure in our sample. In the
course of our analysis we also critically examine the criteria used in
the literature in determining whether systems are fossil or not, and
compare them to the original definition of Jones \rm{et al.} (2003).

The paper is organised out as follows. In Section 2 we describe the
sample selection, observations, data reductions and supplementary
data. Section 3 details our methods of analysis.  In Section 4 our
results are presented and discussed. In Section 5 we summarise our
results and discuss issues arising from them. Our conclusions are
presented in Section 6.\\

Unless otherwise stated, all data presented in this work are scaled to
a cosmology with Hubble constant of 70~km~s$^{-1}$~Mpc$^{-1}$,
$\Omega_M$=0.3 and $\Omega_{\Lambda}$=0.7.

\section{Sample selection, observations and data reductions}
\subsection{Sample selection}
\label{ss}
Five of the groups reported in this paper were selected from the SDSS
maxBCG\footnote{Based on DR6 of the SDSS} catalogue (Koester \rm{et
  al.} 2007). We shall refer to these groups as the ``SDSS sample''.

The maxBCG algorithm identifies clusters as overdensities of red
galaxies.  In the construction of the catalogue, the richness of each
cluster is initially estimated as N$_{gal}$, which is approximately
the number of red sequence galaxies within $\pm$2$\sigma$ of the g -- r
color of the BCG within a fixed 1 h$^{-1}$~Mpc aperture.  The initial
richness estimate (N$_{gal}$) is then used to derive a scaled aperture
(Hansen et al. 2005) and the remeasured richness is called N$_{200}$.
Note, however, that although N$_{200}$ is an effective tracer of mass
(e.g. Johnston et al. 2007; Becker et al. 2007; Rykoff et al. 2008;
Rozo et al. 2009), it is not actually a measure of the number of
galaxies within R$_{200}$, since the values of R$_{200}$ were not
defined in these works.

Now, given that the definition of fossil groups involves the gap
between first and second ranked galaxies within 0.5R$_{200}$, it is
clearly crucial that, for this work, we have an accurate measure of
R$_{200}$.  For the purposes of our study we therefore adopt the value
of R$_{200}$ obtained using the weak lensing analysis of Johnston et
al. (2007) and Sheldon et al. (2009).  In these studies, the maxBCG
systems were stacked in bins of N$_{200}$ (as defined above) in order
to measure M$_{200}$ and R$_{200}$.  Throughout this work we refer to
the R$_{200}$ radius derived in this way as R$_{200,S}$.


The sample presented here was selected from low richness clusters
(9$<$N$_{200}<$25; see Miller \rm{et al.} 2011) and was required to
exhibit an i band magnitude gap of $\geq$2 mag between first and
second ranked galaxies within 0.5R$_{200}$ (as defined above) of the
group centre. The brightest group/cluster galaxy (BCG) was required to be
brighter than $9\times10^{10}~L_{\odot}$ (with the luminosity data
k-corrected to z=0.25) and in the redshift range
0.1~$<$~z~$<$~0.15. Groups whose BCG exhibited evidence for a bright
AGN at the core were excluded in order to maxims the utility of the
low spatial resolution XMM-Newton follow-up that is part of the
programme.

Within each group individual galaxies were then priorities for
spectroscopic observation on the Magellan Baade telescope.
Prioritisation was performed by preferentially selecting galaxies
within 500~kpc of the BCG and brighter than 20~mag in the r
band. Despite the preference for galaxies close to the BGG, candidates
were selected out to the full extent of the IMACS field-of-view
($\sim$30~arcmin or typically $\sim$ 4~Mpc). No galaxies fainter than
21~mag in r were selected. Galaxies with g~--~i colour 0.1~mag redder
than the red sequence identified in the maxBCG catalogue were also
rejected as likely background galaxies.

A total of $\sim$90 galaxies were selected in this way for each group,
requiring two pointings (masks) per group. The success of the
selection scheme is evidenced by the relatively high fraction of
galaxies ($\geq$50\%) that we confirm to be at the redshift of the
central galaxy. However, it should be noted that the scheme results in
a sample that is neither photometrically nor spatially complete.\\

We also report new Gemini GMOS data for the fossil groups RX
J1256.0+2556 and RX J1331.5+1108. Pre-imaging of the groups in g and
i bands was carried out on 2006 February 2 and 2005 February 19,
respectively. Imaging of each group consisted of 3~$\times$~290s
exposures in each waveband.  Calibration to the SDSS photometric
system was carried out using 4 stars in the Landolt
(1992) field PG1323-086. Spectroscopic candidates were selected on
the basis of their apparent magnitudes (m$_i<21.5$~mag) and their
position on the colour-magnitude diagram constructed using galaxies
in the vicinity of the group (i.e. only galaxies close to, or bluer
than, the red sequence visible in the colour-magnitude relation were
selected).  A total of 38 and 22 galaxies were selected in this way
for RX J1256.0+2556 and RX J1331.5+1108, respectively. 

We supplement the above samples with three other fossil groups that
have been spectroscopically studied using GMOS as above to depths
permitting the identification of 20--40 confirmed members:
RX~J1340.6+4018, RX~J1416.4+2315 and RX~J1552.2+2013. These have been
previously reported in the literature by Mendes de Oliveira et
al. 2009; Cypriano, Mendes de Oliveira \& Sodr\'{e}, 2006 and Mendes
de Oliveira, Cypriano \& Sodr\'{e}, 2006, respectively. We shall refer
these five groups (RX J1256.0+2556, RX J1331.5+1108, RX~J1340.6+4018,
RX~J1416.4+2315 and RX~J1552.2+2013) as the ``RXJ sample''\\

For each group we also include, when available, spectroscopic SDSS
data for the observed fields to augment both the literature and new
groups. These relatively bright galaxies, which often include the
central group galaxies, were generally avoided from our Magellan and
Gemini observing plans in order to maximise the number of new
group members identified.\\

It is important to note that the selection criteria for the two
samples (SDSS and RXJ) differed.  The SDSS sample was selected (as
described above) to possess bright central galaxies in low richness
groups, as well as meeting the magnitude gap criterion. The selection
criteria of the RXJ sample are a little less well defined, being
selected (according to Jones \rm{et al.} 2003) by a ``\emph{variety of
  indicators}''. The selection included only high X-ray luminosity
groups with appropriate magnitude gaps and paying ``\emph{particular
  attention}'' to groups with low ratios of X-ray to BCG optical
luminosities and selecting ``\emph{..... system[s] dominated by a
  single galaxy}''. These selection criteria were nominally designed
to reflect the properties of the prototypical fossil group
(RX~J1340.6+4018) first reported by Ponman \rm{et al.} (1994).  We shall
consider the impact of these differing selection criteria in a later
section.

\subsection{Spectroscopic Observations}
The Magellan Baade telescope multi-object spectroscopy of five
candidate fossil groups selected from the maxBCG catalogue was carried
out on the f/2 camera of the IMACS instrument in 2009 Feb. The 300
lines/mm grating was used in conjunction with the ``Spectroscopic 2''
filter in order to maximise the number of spectra that could be fit
onto the CCD. A slit width of 1.0~arcsec was used for all
galaxies. The resultant spectra covered the 4800--8000~\AA\ spectral
range at a resolution of $\sim$6.5~\AA\ and a dispersion of 2.6~\AA/pix
(with $\times$2 spectral binning). The $\sim$30~arcmin field-of-view
results in a spatial extent of $>$4~Mpc at the redshifts of these
groups. Two 1800~s observations of two masks were carried out for each
group.\\

The Gemini GMOS spectroscopic observations of RX~J1256.0+2556 were
carried out on Gemini North on 2006 June 24 (program ID GN-2006A-Q-31).
Observations of RX~J1331.5+1108 were carried out on
Gemini North on 2005 March 7 (GN-2005A-Q-38). Observations
were carried out using the R400 grating and slits of 1 arcsec width,
giving a resolution of 6.5~\AA\ over the 4000 to 8000~\AA. Three
exposures of 2400~s were performed. It should be noted that the
field-of-view of the GMOS instrument ($\sim$5.5$\times$5.5~arcmin) is
considerably smaller than the IMACS instrument on Magellan, resulting
in a spatial extents of 1.2~Mpc and 0.5~Mpc at the redshift of
RX~J1256.0+2556 and RX~J1331.5+1108, respectively.

\subsection{Data reduction}
The Magellan data (systems with prefix SDSS in Table \ref{lxtx}) were
reduced using the {\small COSMOS} pipeline provided by the Magellan
consortium. However, during the analysis it was discovered that the
optical map embedded in the software had not been updated after a
change in the CCD configuration. This resulted in step functions in
the spectral and spatial maps generated by the software. The problems
in the spatial mapping are of no concern for the present work as, for
our purposes, they are adequately handled by the pipeline. However, in
order to compensate for the spectral distortions, it was found
necessary to re-position the data on the CCDs.  This process is only
accurate to about 0.5~pixels ($\sim$50~km s$^{-1}$).  We therefore
assume this value as a minimum error in individual recession velocity
measures. Thereafter, reductions followed a standard procedure of
de-biasing, flat-fielding, wavelength calibration (using Cu-Ar
comparison-lamp exposures), sky-subtraction, cosmic-ray removal and
extraction using the COSMOS pipeline\footnote{http:/obs.carnegiescience.edu/Code/cosmos/Cookbook.html}.\\

Data reductions of the Gemini spectroscopic data were carried out
using the {\small IRAF} Gemini package {\small GMOS} as described in
Mendes de Oliveira \rm{et al.} (2009). Wavelength calibrations were carried
out using Cu-Ar comparison-lamp exposures. Positions and magnitudes
were obtained for all objects using the SExtractor program of Bertin
\& Arnouts (1996).\\

\subsubsection{Measurement of recession velocities}
For the Magellan data, recession velocities were measured using the
Fourier cross-correlation routine ({\small fxcor}) within {\small
  IRAF}. As no velocity standards were observed, a synthetic spectrum
of a typical early-type galaxy was used as a template. In order to
facilitate the identification of group members, the template was first
redshifted by the value of the redshift of the central galaxy of the
group in question, as given by the SDSS spectroscopic survey. All
measured values of recession velocity are therefore with respect to
the central galaxy\footnote{We note that the 1+z
cosmological factor required in the calculation of velocity
dispersion at high redshift is automatically accounted for in this
approach.}.

Only cross-correlations with unambiguous peaks were accepted as valid
measures. However, inspection of the spectra also revealed a number of
galaxies for which unambiguous recession velocities could not be
derived using {\small fxcor}, but that exhibited strong emission
lines. Recession velocities for these galaxies were measured by
fitting a Gaussian profile to the H$_{\alpha}$ emission line. Typical
errors for both absorption and emission line errors were $\sim$75 km
s$^{-1}$.\\

Recession velocity measurements of the Gemini data were performed
using the cross-correlation technique implemented in the {\small RVSAO} 
package within {\small IRAF}. Several galaxy templates were
employed in this analysis with results taken from the template with
the strongest cross-correlation peak. Recession velocities were then
converted to the rest frame of the central galaxy using:

\begin{equation}
V_i = \frac{cz_i-cz_0}{1+z_0},
\end{equation}

\noindent where V$_i$ is the recession velocity of the {\it i}th
galaxy with respect to the BCG, which has redshift z$_0$.\\

In order to maximise the number of \emph{new} cluster members in our
sample we selected against galaxies with pre-measured recession
velocities in the SDSS spectroscopic survey. However, in order to
check the consistency of the two data sets, we did observe 8 galaxies
which were also observed in the SDSS spectroscopic survey (5 in the
Magellan sample galaxies and 3 in the Gemini sample of J1256). A
comparison of the derived values for these galaxies showed our values
to be offset from the SDSS values by --94~$\pm$~35 km/sec and
--59~$\pm$~35 km/sec in the Magellan and Gemini samples respectively,
giving --81$\pm$~35 km/sec for the combined sample. We therefore
offset our data by --81~km/sec before combining our data with that of
the SDSS.\\ 

\section{Supplementary data}
As well as the spectroscopic data detailed above, several sources of
supplementary data were employed in our analysis.

\subsection{X-ray data}
\label{xdata}
The X-ray data for the five new fossil group candidates presented in
this work (groups with prefix SDSS) are taken from Miller \rm{et al.}
(2011). The X-ray data for the remainder of the fossil groups included
in this work were taken from KPJ07.  The data are shown in Table
\ref{xray_data}\footnote{Full galaxy identifiers are given here and
  shown in Table \ref{xray_data}, but throughout the remainder of this
  paper we shall refer to them by abbreviated identifiers in the text
  (e.g. J0906, J1256, etc).}. The Miller \rm{et al.} (2011) data are derived
from Chandra ACIS--S3 snapshots, while the KPJ07 data are derived from
deeper Chandra ACIS--S3 observations. We note that the upper-limit of
log(L$_X$)$<$43.46 erg s$^{-1}$ for the non-detection of J0906 is
still well above the X-ray luminosity criterion for fossil groups
(log(L$_X$)$>$41.7 erg s$^{-1}$). This group \emph{may} still
therefore meet this criterion, and its eligibility as a fossil
candidate will be reviewed in a later section.

Table \ref{xray_data} also shows values of R$_{200}$ (which we take in
this work to be an approximation for the virial radius). Values of
R$_{200}$ are required in both the selection process (in order to
determine the luminosity gap between the first and second ranked
galaxies within 0.5R$_{200}$), as well as in the later analysis.

Table \ref{xray_data} presents both the values of R$_{200}$ used in
the selection process (R$_{200,S}$) and those derived form the X-ray
temperatures also presented in the table (R$_{200,X}$), which
were derived using the full cosmological form of the expression
given in Helsdon \& Ponman (2003):

\begin{equation}
R_{200,X} = 1.14\sqrt{T_X}~h_{50}^{-1}(z)Mpc,
\label{xrad}
\end{equation}

\noindent where $h_{50}(z) = h_{50}(\Omega_M(1+z)^3+\Omega_{\Lambda})^{0.5}$ 
assumes a $\Omega=1$ universe and the 1.14 coefficient was derived
from the results of the N-body/SPH simulations of Navarro, Frenk \&
White (1995). The values of R$_{200,X}$ so derived are given in Table
\ref{xray_data}. However, J0906 is a non-detection. Consequently, only
an upper limit on its X-ray luminosity could be estimated and its
X-ray temperature is unconstrained. We have therefore assumed a value
of 1~Mpc for R$_{200,X}$ of this group based on on its dynamical
properties (see Section \ref{dyn_prop}).

An estimate of the masses of the groups can be made directly from the
values of R$_{200,X}$ using:

\begin{equation}
M_{200,X} = \frac{4}{3} \pi R_{200,X}^{3}.200.\rho_{crit}(z)
\label{xmass}
\end{equation}

Where $\rho_{crit}(z)$ is the critical density at redshift z. The
values of M$_{200,X}$ so derived are given in Table \ref{xray_data}.
Table \ref{xray_data} shows clear discrepancies between the values of
R$_{200,S}$ and R$_{200,X}$. 

For the RXJ sample (but with the exception of J1340) R$_{200,S}$ was
based on relationships between L$_X$ and kT, and kT and R$_{200}$ (see
Jones \rm{et al.} 2003 for details). However, for J1256, J1416 and J1552,
the ROSAT X-ray data, upon which these estimates were based, yielded
X-ray luminosities (and therefore R$_{200}$ values) significantly
lower than the subsequent, higher resolution Chandra data presented in
KPJ07. In the case of J1340, no value of R$_{200}$ was quoted in the
original Ponman \rm{et al.} (1994) paper. We therefore estimated
R$_{200,S}$ using the ROSAT X-ray temperature given in KPJ07 and the
equation given in Jones \rm{et al.} (2003)

In the case of the SDSS sample R$_{200,S}$ values were based on the
results of the weak lensing analysis of Johnston \rm{et al.} 2007) for which
groups were stacked by their richness (N$_{gal}$; see Section \ref{ss}).
We shall discuss the cause of the discrepancies between R$_{200,S}$
and R$_{200,X}$ values in this sample in a later section.

Unless otherwise specified, throughout this work virial radii are
taken to be the R$_{200,X}$ values given in Table \ref{xray_data}.\\

\begin{table*}
\begin{centering}
\caption{For groups with prefix SDSS (the SDSS sample), X-ray data are
  the Chandra data of Miller \rm{et al.} (2011). For the systems in the RXJ
  sample the Chandra data of KPJ07 are presented. X-ray luminosities
  and temperatures are specified for an aperture equal in size to
  R$_{200}$. The values of R$_{200}$ used in the selection of the
  target systems (R$_{200,S}$) are shown (see text). Values derived
  from the X-ray data (R$_{200,X}$) are also presented.  For SDSS
  J0906+0301, which was undetected in the X-ray, the R$_{200,X}$ is
  assumed to be 1.0~Mpc based on its dynamical properties (See Section
  \ref{dyn_prop}). In all other cases R$_{200,X}$ was calculated from
  the X-ray temperature using Equation \ref{xrad}. Group masses
  derived using Eqn. \ref{xmass} are also presented.}
\begin{tabular}{|l|c|r|c|c|c|c|}
\hline
Group  & z &  log(L$_X$)  &  kT$_X$ &R$_{200,S}$ & R$_{200,X}$ & log(M$_{200,X}$)\\
       &   &(erg~s$^{-1}$) & (keV)  &(Mpc)&(Mpc)& (M$_\odot$) \\
\hline
SDSS J0906+0301 & 0.1359  &$<$43.29 &  --   & 0.66 & (1.0)$^*$& (14.12 $\pm$ 0.25)\\
SDSS J1045+0420 & 0.1539  & 44.01   &  2.47 & 0.76 & 1.19& 14.35 $\pm$ 0.14\\
SDSS J1136+0713 & 0.1030  & 43.59   &  2.64 & 0.86 & 1.26& 14.41 $\pm$ 0.23\\
SDSS J0856+0553 & 0.0939  & 43.92   &  2.73 & 0.83 & 1.29& 14.43 $\pm$ 0.19\\
SDSS J1017+0156 & 0.1177  & 42.99   &  2.13 & 0.74 & 1.12& 14.26 $\pm$ 0.24\\
\hline
RX J1256.0+2556 & 0.2327  & 43.70  &  2.63  & 0.69 & 1.18& 14.38 $\pm$ 0.20\\
RX J1331.5+1108 & 0.0802  & 42.32  &  0.81  & 0.71 & 0.71& 13.65 $\pm$ 0.10\\
RX J1340.5+4017 & 0.1719  & 42.72  &  1.16  & 0.75 & 0.81& 13.86 $\pm$ 0.23\\
RX J1416.4+2315 & 0.1381  & 44.23  &  4.00  & 0.93 & 1.52& 14.66 $\pm$ 0.10\\
RX J1552.2+2013 & 0.1357  & 43.78  &  2.85  & 0.83 & 1.29& 14.45 $\pm$ 0.14\\
\hline
\end{tabular}
\label{xray_data}
\\
$^*$Assumed value.
\end{centering}
\end{table*}

\begin{figure}
\centerline{\psfig{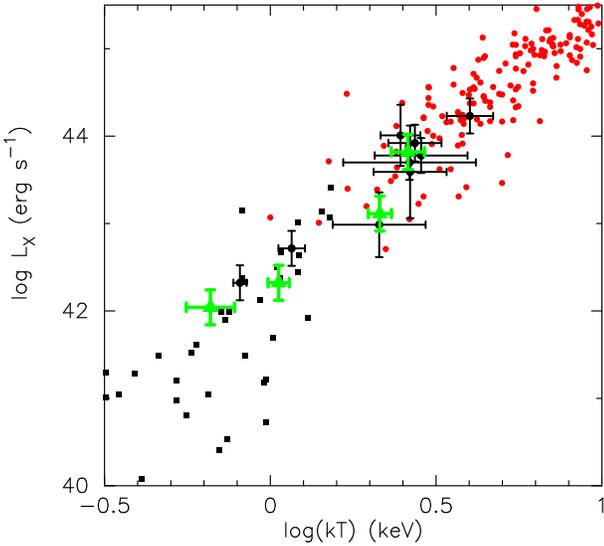}}
\caption{Fossil groups are plotted in the L$_X$--T$_X$ plane and
  compared to literature values for ``normal'' systems.  Literature
  values for normal clusters (Wu, Xue \& Fang, 1999) are shown as red
  dots, while literature values for normal groups (Osmond \& Ponman
  2004) are shown as black squares. Fossils taken from KPJ07 are shown as
  green squares. The fossil groups analysed in this work are shown
  as black dots (with error bars).}
\label{lxtx}
\end{figure}

In Fig. \ref{lxtx} we plot a comparison of the L$_X$--T$_X$ relation
for our fossil groups to the literature relation for ``normal''
systems. In this figure normal groups are shown as black dots (Osmond
\& Ponman, 2004) and normal clusters as red dots (Wu et al.
1999). The figure shows the fossil groups to be generally consistent
with normal systems. However, we note that, as found in many previous
studies (e.g. Mendes de Oliveira \rm{et al.} 2006 and 2009; Cypriano et
al. 2006; KPJ07), a significant number of the fossil groups exhibit
values consistent with \emph{clusters} rather than groups. However,
there are four systems with group-like X-ray
properties. Interestingly, these appear to lie above the L$_X$--T$_X$
relation for normal groups. We shall comment further on these trends
in a later section.\\

\subsection{Control samples from the literature}
\label{control}
Our analysis involve the comparison of our fossil group candidates
with ``normal'' systems.  The sources for normal groups were taken
from the GEMS project of Osmond \& Ponman (2004), supplemented by
groups from the study of Girardi \rm{et al.}  (2002). For clusters,
X-ray luminosities were taken from Wu \rm{et al.} (1999) and Zhang
\rm{et al.} (2011). X-ray temperatures were taken from Wu \rm{et al.} (1999)
and r band luminosities from Girardi \rm{et al.} (2002).
However, we note that, these studies were not specifically
selected to possess low m$_{12}$ values. The samples \emph{likely}
therefore contain a few systems that we would consider fossil systems.

When converting galaxy luminosities from the literature to the r band
used in this work values of B~--~R, B~--~r and r~--~i were taken to be
1.57, 1.33 and 0.4 mag, respectively. To convert absolute magnitudes
into solar luminosities, values of the solar B, R, r and i band
absolute magnitudes were taken to be 5.48, 4.42, 4.76 and 4.58 mag,
respectively. Finally, the value of B$_j$~--~B in galaxies was taken
to be the solar value when converting the Girardi \rm{et al.} (2002)
luminosities to Sloan r band. \\

All literature values were converted to the H$_0$=70 km s$^{-1}$
Mpc$^{-1}$, $\Omega_M$=0.3, $\Omega_{\Lambda}$=0.7 cosmology used
throughout this paper.\\

\section{Analysis}
In this section we detail each of the elements of our analyses of the
spectroscopic and photometric data. The data are presented in Appendix B.\\

\subsection{Group velocities and velocity dispersions}
In this section we detail our estimates of group velocities, velocity
dispersions and dynamical virial radii (i.e. R$_{200,dyn}$) of the fossil
groups.

The average velocity of the group was also calculated as:

\begin{equation}
RV_{group} =\frac{\Sigma V_i}{N} \pm  \frac{\sigma_{200}}{\sqrt{N}}~km s^{-1},
\label{delrv}
\end{equation}

\noindent where $V_i$ is the recession velocity of the $i$th galaxy
within R$_{200,X}$ and N is the total number of non-BCG galaxies within
R$_{200,X}$ (the BCG is excluded from this calculation).  The offset of
the BCG with respect to the group average
($\Delta$RV$_{BCG}$=RV$_{BCG}$-RV$_{group}$) was then calculated
(Table \ref{data}).  The velocity dispersion ($\sigma_{200}$) of each
group was estimated from the recession velocities by:\\

\begin{equation}
\sigma_{200} = \sqrt{\frac{\Sigma (V_i-RV_{group})^2}{N-1}} \pm  \frac{\sigma_{200}}{\sqrt{2(N-1)}}~km s^{-1}.
\label{vdisp}
\end{equation}

Both of the above definitions require an estimate of R$_{200}$. For
these estimates, and generally throughout this work, we use the values
derived from the X-ray temperature as described in Section \ref{xdata}
and given in Table \ref{xray_data}.  However, for comparison purposes
we also make virial radius estimates based on the observed velocity
dispersion (which we shall refer to as \emph{dynamical} virial radius).

\subsection{Dynamical virial radii}
Dynamical virial radii can be estimated from kinematic data using
expressions which express the virial radius as being proportional to
the velocity dispersion.  From the virial theorem, Carlberg \rm{et al.}
(1997) derived an expression for R$_{200}$:

\begin{equation}
R_{200,dyn} = \frac{\sqrt{3}\sigma_{200}}{10H(z)}~Mpc,
\label{r200}
\end{equation}

\noindent where $\sigma_{200}$ is the velocity dispersion of galaxies
within R$_{200,X}$ as defined above, and H(z) is the Hubble constant at
redshift of the group.  Alternatively, Girardi et al (1998) use both
virial theory \emph{and} observational data to derive an expression
for the virial radius (at unstated over-density):

\begin{equation}
R_{vir,dyn} = \frac{0.2\sigma_{200}}{H_0}~Mpc.
\label{rvir}
\end{equation}

\noindent Both of the above equations are directly proportional
to the velocity dispersion, and differ only in the constants of
proportionality  (with the Girardi et al (1998) values larger by a
factor of $\sim$15\%). We therefore use Equation \ref{r200} above for
estimates of the dynamical virial radii and leave it to the reader to
apply the $\sim$15\% offset if the Girardi et al (1998) values are required.\\

\subsection{Dynamical masses}
We also make estimates of the dynamical mass within R$_{200,X}$ of each group or cluster 
using our dynamical data. These are calculated using the expression  (Ramella \rm{et al.} 2004):

\begin{equation}
M_{200,dyn} = \frac{3}{G}\sigma_{200}^2R_{200,X},
\label{mass}
\end{equation}

\noindent which can be expressed in the more convenient form:

\begin{equation}
M_{200,dyn} = 6.975\left(\frac{\sigma_{200}}{1000~km s^{-1}}\right)^2\left(\frac{R_{200,X}}{1~Mpc}\right) \times10^{14}~M_{\odot},
\label{mass1}
\end{equation}

\noindent where R$_{200,X}$ is calculated using Equation \ref{xrad}.

\subsection{Composite luminosity function}
The spectroscopy for the fossil candidates in this paper
obtained using the Magellan telescope (i.e. the SDSS sample) covers a
significant radial extent in each system (i.e $>$R$_{200,X}$). This
allows the accurate determination of the luminosity function (LF) of
galaxies within R$_{200,X}$ in these systems. Although a few
determinations of the luminosity function of individual fossil groups
has been attempted in the literature (Mendes de Oliveira \rm{et al.} 2006
and 2009, for J1552 and J1340 and Cypriano \rm{et al.} 2006 for J1416)
these were within radii smaller than R$_{200,X}$.  This is
therefore the first determination of the LF of fossil groups which
include more than 30 galaxies per group and reach out to R$_{200,X}$.

To calculate the LF of each group, we considered all galaxies inside a
projected radius corresponding to R$_{200,X}$ of
the group. This requires the determination of the selection function
$S(m')$ in each group in order to estimate the completeness of the
spectroscopy.  This was done using the following equation:

\begin{equation}
S(m') = \frac{\#GAL_z(m')}{\#GAL(m')}\,\,,
\end{equation}

\noindent where $\#GAL_z(m')$ is the total number of galaxies with
known spectroscopic redshifts, being member galaxies or not, and
$\#GAL(m')$ is the total number of galaxies in the region as
identified via photometry (from SDSS), in both cases for galaxies with
magnitude $m$ such that $\mid m$ - $m'\mid$ $<$ $\Delta m$. Membership
to the group was defined in the velocity range within 2000 km/s from
the velocity of the central galaxy and within R$_{200,X}$ of the
position of the central galaxy. Then the LF is defined by:

\begin{equation}
LF(m') = \frac{\#GAL_{z,grp}(m')}{S(m')}\,\,,
\end{equation}

\noindent where $\#GAL_{z,grp}(m')$ is the number of member galaxies
as determined by spectroscopy.

Thus, the individual LF for each group in a given band was obtained
by simply dividing the number of galaxies in each bin of absolute
magnitude by the completeness fraction.  The second step was then to
construct the composite LF by averaging, bin per bin, the individual LFs of
each of the five groups for each band g, r, and i.

Finally the galaxy distributions were fitted by the Schechter function
(Schechter 1976).

\subsection{Total optical luminosities}
Part of our analysis considers the total optical luminosities of the
groups.

For the SDSS sample, which was well sampled out to R$_{200}$, and for
which the selection functions and completeness were estimated during
the construction of the composite luminosity function, we estimate the
total optical luminosities using completeness-corrected data.

For the RXJ sample a more complex procedure was adopted, since
composite luminosity functions were not available and our
spectroscopic data are only well sampled within
~0.7R$_{200}$. However, our use of the data from the SDSS database
means that the both RXJ and SDSS samples are 100\% complete all the
way out to R$_{200}$ down to an apparent magnitude of 17.7 (an
absolute magnitude of approximately --21 at the redshift of our
groups).  We therefore use the SDSS sampĺe to estimated the effect on
the total luminosity of the poor spatial sampling and completeness in
the RXJ sample.

We begin by simply adding the optical luminosities of all
the identified group members. Now, in the SDSS groups, $\sim$20\% of
the galaxies below an apparent magnitude of 17.7 lie between
0.7R$_{200}$ and R$_{200}$. For each RXJ group, we therefore added to the
total luminosity 25\% of the light in galaxies that are fainter than
this limit and lie within 0.7R$_{200}$. This increases the logarithmic
luminosity estimates by $\sim$0.03~dex (with a range from 0.01 to
0.04~dex). We then assumed that the completeness of each group in the
RXJ sample was the same as the average of the SDSS sample. This
further increases the luminosity estimates by $\sim$0.07~dex (with a
range of 0.04 to 0.1~dex). 

\section{Results}
In this section we detail the results of our analysis. The recession
velocities and apparent magnitudes of all new data presented in this
paper are given in Appendix B. We begin by considering the results of
our kinematic analysis.

\begin{figure*}
\centerline{\psfig{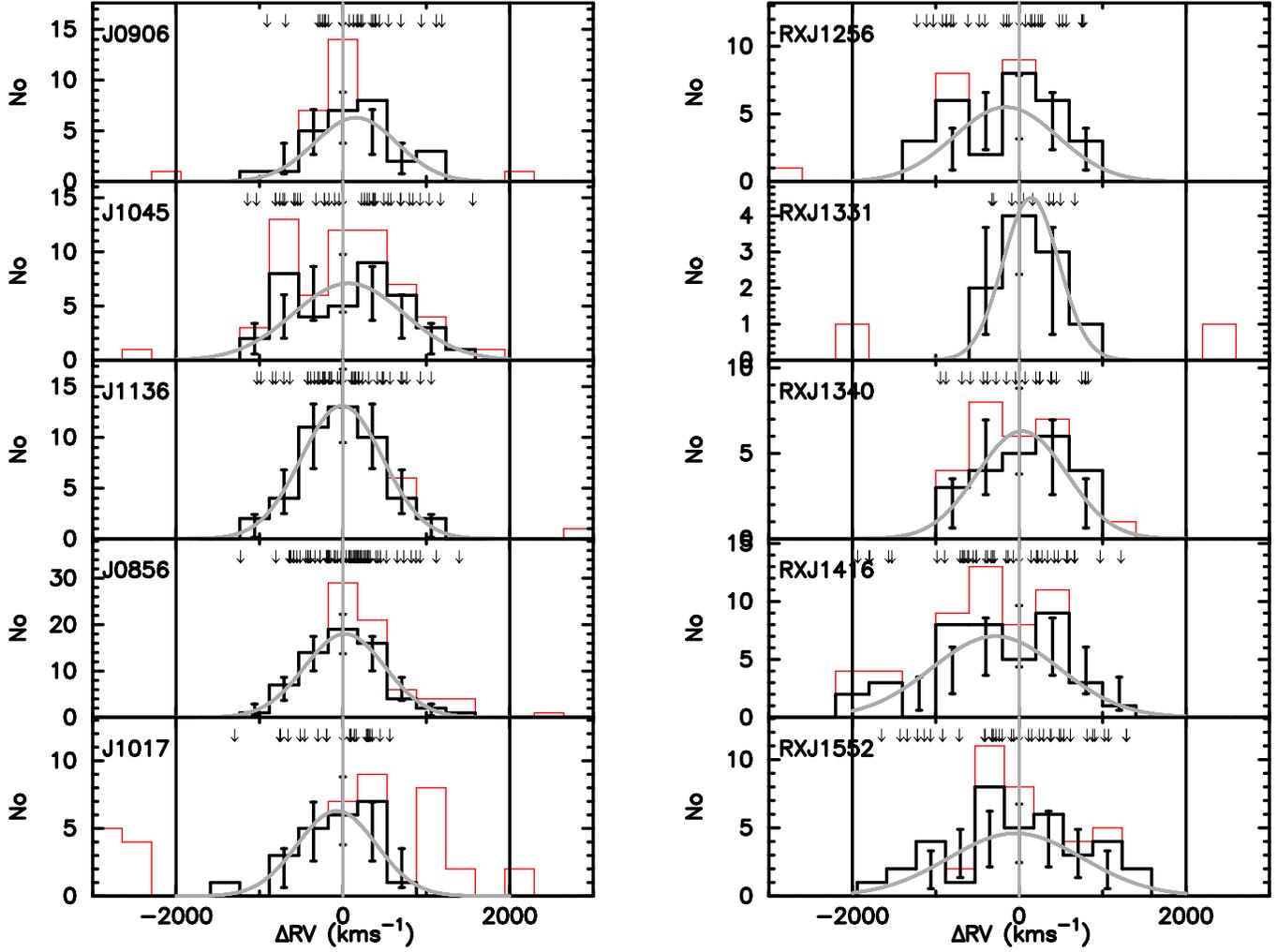}}
\caption{Distributions of recession velocities about the average group
  velocity. Arrows mark the recession velocities of individual
  galaxies. Vertical black lines represent the 2000 km s$^{-1}$
  velocity limit of group members applied to all groups.  Vertical
  grey lines represent the velocities of the BCG in each group
  ($\Delta$RV$_{BCG}$ ; Table \ref{data}). Histograms in black
  include all member galaxies within the R$_{200,X}$. Red histograms
  show galaxies outside the velocity or radius limits. Gaussian
  distributions matching the recession velocity distributions (i.e
  assuming the velocity dispersions listed in Table \ref{data}) are
  shown as grey lines with Poisson errors to aid in assessing the
  significance of apparent velocity substructure. Significant
  asymmetries and/or discontinuities are visible in the distributions
  of J1017, J1256 and J1416.}
\label{rvhistos}
\end{figure*}

\subsection{Recession velocities}
\label{rvs}
The distributions of the recession velocities of galaxies within
$\sim$3~Mpc of the central galaxy are shown in Fig. \ref{rvhistos}.
This figure shows that the majority of the groups exhibit recession
velocity distributions that are clearly delineated, and near symmetric
about zero velocity (i.e. the average velocity of the group). However,
there are indications of skew distributions and gaps in the recession
velocity distributions in a few cases. Other possible signs of
disturbance or background contamination that were considered were
large offsets in BCG velocity (Table \ref{data}), and spatial groupings
of galaxies with similar recession velocities, In order to measure
gaps and skewness, we preformed an analysis using the {\small ROSTAT}
software of Beers, Flynn \& Gebhardt \rm{et al.} (1990). The results
of this analysis were then combined with the BCG velocity offsets and
a visual inspection for spatial groupings. In three cases (J1017,
J1256 and J1416) the groups exhibited positive signs from three of the
four criteria listed above. We therefore took these groups as being
the most likely to be either contaminated by foreground/background
structures or out of equilibrium (i.e. are unvirialised). These groups
were therefore used to test the possible affects of these apparent
irregularities on our derived velocity dispersions and dynamical
virial radii.

\begin{table*}
\begin{centering}
\caption{Dynamical data derived within R$_{200,X}$. The table shows n --
  the number of galaxies within R$_{200,X}$ (with the number of galaxies
  taken from the SDSS in brackets), the estimated velocity
  dispersions and the offset of the BGG velocity with respect to the
  group average ($\Delta$RV$_{BCG}$). R$_{200,dyn}$ and masses estimated
  from Equations \ref{r200} and \ref{mass} are also presented. Errors
  in $\sigma$ and $\Delta$RV$_{BCG}$ were calculated according to
  Equations \ref{vdisp} and \ref{delrv} and were propagated through
  Equations \ref{r200} and \ref{mass} for R$_{200,dyn}$ and M$_{200,dyn}$.}
\begin{tabular}{|l|c|c|c|c|c|}  
\hline
Group    &   n    &$\sigma$& $\Delta$RV$_{BCG}$ &R$_{200,dyn}$& log(M$_{200,dyn}$)\\
&& (km s$^{-1}$) & (km s$^{-1}$)   &  (Mpc)&(M$_{\odot}$)\\
\hline
SDSS J0906    &  25(1)    & 506$\pm$72 & -154$\pm$103 &   1.17$\pm$0.16&14.25$\pm$0.21\\
SDSS J1045    &  38(2)    & 664$\pm$77 &  -69$\pm$109&    1.52$\pm$0.18&14.58$\pm$0.13\\ 
SDSS J1136    &  45(2)    & 490$\pm$52 &   11$\pm$74 &    1.15$\pm$0.12&14.30$\pm$0.22\\
SDSS J0856    &  63(17)    & 478$\pm$43 &  -24$\pm$61 &    1.13$\pm$0.11&14.26$\pm$0.16\\
SDSS J1017    &  23(1)    & 474$\pm$71 &   73$\pm$101&    1.11$\pm$0.17&14.23$\pm$0.29\\
RXJ J1256    &  28(1)    & 622$\pm$84 &  159$\pm$120&    1.37$\pm$0.19&14.50$\pm$0.40\\
RXJ J1331    &  10(6)    & 338$\pm$77 & -142$\pm$111&    0.80$\pm$0.18&13.74$\pm$0.25\\ 
RXJ J1340    &  22(2)    & 537$\pm$82 &  -34$\pm$117&    1.22$\pm$0.19&14.21$\pm$0.10\\
RXJ J1416    &  40(11)    & 815$\pm$87 &  285$\pm$124&    1.89$\pm$0.20&14.85$\pm$0.15\\
RXJ J1552    &  35(8)    & 803$\pm$96 &   43$\pm$138&    1.86$\pm$0.23&14.76$\pm$0.28\\
\hline
\end{tabular}
\label{data}
\end{centering}
\end{table*}

The analysis is presented in detail in Appendix \ref{app}. In brief,
while hints of substructure in the spatial and kinematic data can be
seen in a number of systems, the relatively low numbers of 
members and incomplete spatial coverage of our data preclude definitive
statements about the dynamical status of these systems.  In our
analysis we therefore simply estimated the magnitudes of the effects
such substructure, if real, might have on our derived parameters for
the three most obvious potential cases. We find that while small
quantitative effects may be present, these do not \emph{qualitatively}
affect our results.  We therefore continue to use the values derived
from all galaxies observed within R$_{200,X}$ throughout this
work. Clearly, follow-up observations of these systems to improve the
spatial coverage and depth of spectroscopically confirmed memberships
are highly desirable.

\subsection{Dynamical properties}
\label{dyn_prop}
The dynamical properties (velocity dispersions, average group
velocities, dynamical R$_{200}$ values  and dynamical masses) are presented in Table
\ref{data}. 

A comparison of our velocity dispersion results with those of KPJ07,
for the four systems common to both studies, shows them to be
consistent, with our results exhibiting an offset and rms of +54 and
104~km~s$^{-1}$ with respect to KPJ07.  These are easily within
1$\sigma$ in all cases. Comparison of the log(mass) estimates within
R$_{200}$ are also consistent with our results exhibiting an offset
and rms of -0.02 and 0.23~dex, respectively. These values are also
consistent with the masses derived solely from the X-ray temperature
given in Table \ref{xray_data}, exhibiting an average offset of only
+0.1~dex (i.e within 1$\sigma$.  This clearly shows that the
relationship between the values of kT$_X$ and velocity dispersion
presented in this work is in general consistent with virial theory.

Comparison of the dynamical R$_{200}$ with the X-ray derived values
given in Table \ref{xray_data} shows good agreement, with the
dynamical values on average 0.21~Mpc (18\%) larger than the X-ray
values with an RMS scatter of 0.20~Mpc (18\%). Use of the Girardi et
al (1998) expression (Equation \ref{rvir}) would have resulted in
values $\sim$15\% larger still.

A striking feature of the dynamical data is the magnitude of the
velocity dispersion and associated mass estimates.  The derived masses
are, in all but one or two cases, greater than
10$^{14}$~M$_{\odot}$. These masses are more consistent with poor
\emph{clusters} than with groups -- in accord with the trend suggested
by the X-ray luminosities and temperatures (Section \ref{xdata}). It
should be noted that the two systems with the lowest masses (J1331 and
J1340) are the systems identified as possessing low X-ray temperatures
and luminosities in Fig. \ref{lxtx}.

The data therefore indicate a consistency between the dynamical and
X-ray properties. We therefore next consider the scaling relations
that relate dynamical, X-ray and optical properties of groups and
clusters.

\subsection{Scaling relations}
\label{scale}

\begin{figure*}
\centerline{\psfig{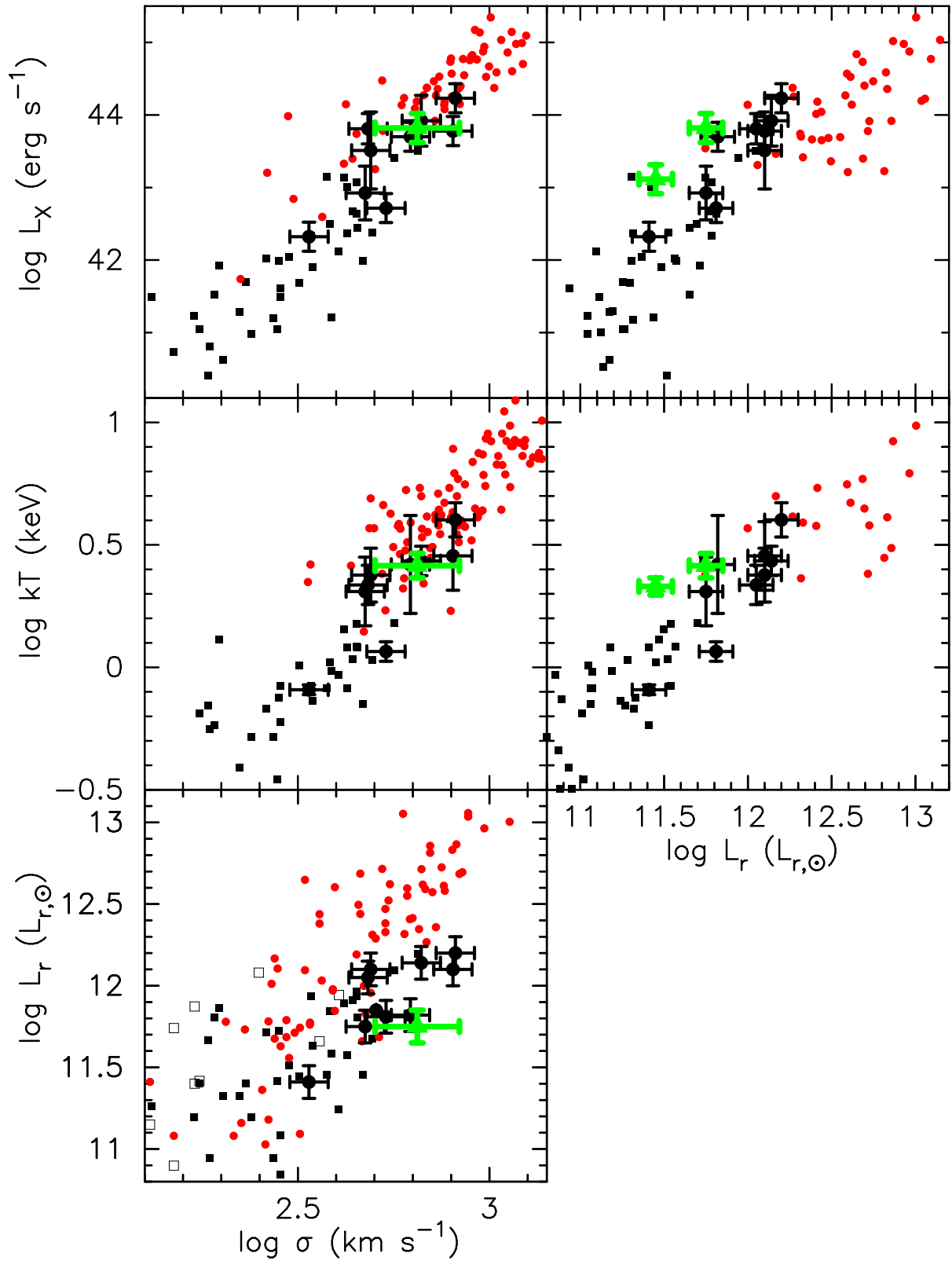}}
\caption{The scaling relations of our sample of fossil groups is
  compared to ``normal'' systems from the literature. The data
  presented in paper are shown as black dots with error bars. Three
  groups from KPJ07 are shown as green dots with error bars. Samples
  of ``normal'' groups are shown as black squares (filled symbols from
  Osmond \& Ponman 2004 and open symbols from Girardi \rm{et al.}
  2002). Clusters are shown as red dots.}
\label{sr}
\end{figure*}

The optical data are presented in Table \ref{photo}, while the scaling
relations are shown in Fig. \ref{sr}. In this figure the data from the
present work are shown as black dots with error bars. We also include
the data for three fossil groups from KPJ07. These are shown as green
dots with error bars.
 
The figure also shows values for normal groups and clusters from the
literature (as detailed in Section \ref{control}). It is worth recalling that
the ``normal'' systems may, in fact, contain some fossil systems, as
these were not expressly excluded during the construction of these
samples.

The trend noted in the L$_X$--T$_X$ plot for the majority of the
fossil groups to be more similar to galaxy \emph{clusters} than groups
is also evident in the plots of L$_X$--$\sigma$ and T$_X$--$\sigma$ of
Fig. \ref{sr}.  We therefore find that all three of these commonly
used proxies for mass are in accord, indicating that the majority of
the systems in our sample possess masses $\sim$10$^{14}$~M$_{\odot}$,
or greater.  We note that the group that was a non-detection in the
X-ray (J0906) has a velocity dispersion of $>$500 km~s$^{-1}$ (log
$\sigma \sim$ 2.7).  For the properties of this system to be
consistent with our other data, we should expect J0906 to possess a
log~L$_X~\sim$~43.0. The upper limit of log~L$_X\sim$~43.3 found for
this group therefore does not preclude this system from either meeting
the fossil group criteria (log~L$_X>$41.7), or following the same
scaling relations as the remainder of our sample.

However, a severe mismatch with cluster data is evident in the
L$_r$--$\sigma$ plot, with the fossil groups exhibiting r band
luminosities $\sim$0.5~dex lower than clusters of the same velocity
dispersion. Examination of the L$_X$--L$_r$ plot of Fig. \ref{sr} (top
right) shows the fossil groups to lie on the outer envelope of the
locus of normal groups. This trend has been noted in previous works
(e.g.  Vikhlinin \rm{et al.}  1999; Jones \rm{et al.} 2003; KPJ07) and
is often interpreted as a X-ray luminosity excess. However, considering the
plots with velocity dispersion, it appears that the data are
more accurately interpreted as an optical luminosity deficit. Indeed, we note
that, if we compensate for the $\sim$0.5~dex deficit in L$_r$
suggested by the L$_r$--$\sigma$ plot, then the fossil groups would
fall in the cluster region of the L$_X$--L$_r$ plot.

The disparity in r band luminosity is harder to discern in the
T$_X$--L$_r$ plot. However, we note that the displacement of the
fossil group data points by the $\sim$0.5~dex suggested by the
L$_r$--$\sigma$ plot leaves most of the fossil groups consistent with
the trends shown by normal cluster-like systems. There are, however, two
notable exceptions - J1331 and J1340 (the two systems with the lowest
X-ray temperatures), which already exhibit luminosities high for their
X-ray temperatures.  We note that these groups were also amongst those
identified in Fig. \ref{lxtx} as exhibiting low T$_X$ for their
L$_X$. It is therefore possible that these two systems represent a
separate, distinct population (i.e. following different scaling
relations) from their more massive counterparts. Clearly, an expansion
of the data set at low masses (low $\sigma$, L$_X$ and T$_X$) is
highly desirable to address this point.

\begin{figure}
\centerline{\psfig{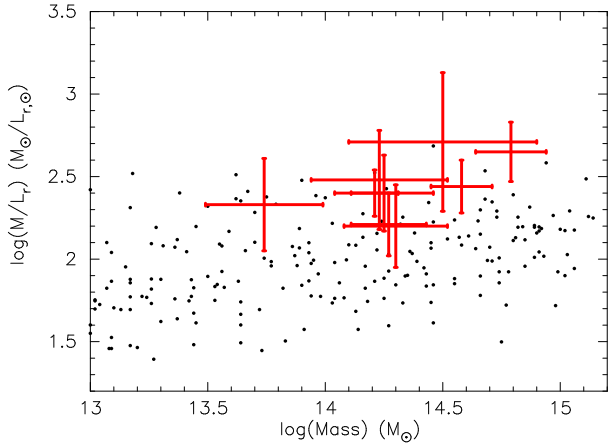}}
\caption{Mass-to-light ratios are plotted against mass (with both
  parameters estimated within R$_{200,X}$). Literature values
for normal groups and clusters (Girardi \rm{et al.} 2002) are shown as black dots.}
\label{m2l}
\end{figure}

The disparity in r band luminosity between the majority of the fossil
groups and normal systems of the same mass indicates that the
mass-to-light ratios of the fossil groups are $\sim$3 times larger
than normal systems of the same mass. This is demonstrated in Table
\ref{photo} and Fig. \ref{m2l}, in which our mass and mass-to-light
values (determined within R$_{200,X}$) are compared to the values for
normal systems from Girardi \rm{et al.} (2002). The Girardi \rm{et
  al.} values were estimated within R$_{vir,dyn}$ (Equation
\ref{rvir}) and are therefore well matched to our data. It is evident
from Fig. \ref{m2l} that the fossil groups lie on or above the highest
mass-to-light ratios exhibited by normal systems. It is also
interesting to note that the two low-mass fossil systems (J1331 and
J1340) also possess high mass-to-light ratios. Therefore, even if
these systems do signal the existence of a distinct low-mass
population, this too would seem to exhibit high mass-to-light ratios,
and their positions in the L$_X$--T$_X$, T$_X$--$\sigma$ and
T$_X$--L$_{r}$ planes would suggest low X-ray temperatures for their masses.

We note that our findings (velocity dispersions, masses and total
luminosities for the five overlapping systems) are in good accord with
KPJ07. However, examination of Fig. 10 of KPJ07 suggest that our
estimates of mass-to-light ratio are not. However, in a recent
re-examination of their data (private communication), the authors of
KPJ07 discovered that the values of mass used in the construction of
their plot were in fact M$_{500}$, rather than the M$_{200}$ required
for the comparsion to the Girardi et al (2002) data. The use of the
appropriate mass values would have resulted in their finding
considerably higher mass-to-light ratios in the fossil group sample
than those in the normal systems representeed by the Girardi et
al. (2002) data.

To summarise, our consideration of the scaling relations of fossil
groups indicate that the most important parameter differentiating the
fossil sample from normal systems is their optical luminosity, with
Fig. \ref{m2l} demonstrating that fossil systems possess mass-to-light
ratios approximately three times that of normal systems of the same
mass. This corresponds to a deficit of \emph{twice} the \emph{total}
observed luminosity (including the bright BCGs), or more than three
times the total luminosity of \emph{all} the non-BCG galaxies. Such
deficits can not be explained simply by the absence of one or two
bright galaxies, neither can they be explained by completeness
issues. This was confirmed by a simple test performed on the SDSS
sample, in which we constructed an alternative completeness function
on the basis that \emph{all} galaxies detected within the virial
radius were considered group members unless specifically excluded by
the spectroscopy. This is clearly produces a gross overestimate of the
total luminosities of these groups. Nevertheless, the increase in
luminosity found of $\sim$0.25~dex does not qualitatively change our
conclusion that fossils exhibit mass-to-light ratios greater than
those of normal systems of the same mass. This can most easily be seen
by considering Fig. \ref{m2l} as the reduction in log[M/L] of
0.25~dex, which is of the same magnitude as the plotted errors, does
not change the conclusion that the fossils lie in the outer envelope
of data for normal systems. The high mass-to-light ratios that we find
in fossil systems is therefore robust to completeness and calibration
issues.

We next consider the optical properties of the sample
in more detail in order to better characterise these systems.

\subsection{Optical properties}
\label{photanal}
In this section we look in detail at the optical properties of our
fossil systems (luminosities, luminosity functions and m$_{12}$ gaps).
The data are presented in Table \ref{photo}. 

\subsubsection{The m$_{12}$ gaps}
The estimated values of m$_{12}$ are presented in Table
\ref{photo}. Two sets of values are given, one (m$_{12,S}$) is
measured within the radius used in the selection process
(R$_{200,S}$), the other (m$_{12,X}$) is measured within
R$_{200,X}$. Now, we demonstrated in Section \ref{xdata} that for our
samples R$_{200,X}$ is typically $\sim$50\% larger than
R$_{200,S}$. It can be seen from Table \ref{photo} that this results
in significant reductions in the observed m$_{12}$ in many cases, with
only 3/10 \emph{strictly} meeting the Jones \rm{et al.} (2003)
criterion of a 2 magnitude gap within 0.5R$_{200}$.

As previously noted (Section \ref{xdata}), in the RXJ sample,
the disparity between R$_{200,S}$ and R$_{200,X}$ is caused by the
difference in X-ray properties derived from the ROSAT data reported in
Jones \rm{et al.} (2003) and the higher quality Chandra data reported
in KPJ07 (and used in this work).

In the SDSS sample, we find that the discrepancy can be explained by
the high mass-to-light ratios found in these groups. Recall that
Johnston \rm{et al.} (2007) binned the maxBCG systems by optical
richness in order to carry out the weak lensing analysis from which
the R$_{200,S}$ values were derived. This implicitly assumes that all
systems of the same richness have the same mass (and hence the same
R$_{200}$). Now, given that we find the systems in our sample of FGs
to have high mass for their luminosity (and richness), it follows that
the values of R$_{200}$ derived for these systems by the Johnston
\rm{et al.} (2007) analysis will be underestimated.

Clearly these issues have a significant effect on the values of
m$_{12}$ derived for our samples. However, this problem is not
confined to the two samples considered in our paper. Indeed, in
\emph{most} studies in the literature the definition of fossil groups
proposed in Jones \rm{et al.} (2003) - i.e.  a 2 magnitude gap within
half the virial radius - is not strictly followed. For instance, in
some studies (e.g.  Santos, Mendes del Oliveira \& Sodr\'{e} 2007;
Smith \rm{et al.} 2011) a fixed radius of 0.5~Mpc is used (in systems
many of which have properties suggesting viral radii greater
1.0~Mpc). In other studies (e.g. Barbera \rm{et al.}  2009; Voevodkin
\rm{et al.}  2010; Aguerri \rm{et al.} 2011) both a radius less than
0.5R$_{200}$ \emph{and} a reduced m$_{12}$ gap were employed as the
definition of FGs. In those studies that \emph{do} strictly apply the
Jones \rm{et al.}  (2003) criteria (e.g. Zibetti, Pierini \& Pratt
2009; D\'{e}mocl\`{e}s \rm{et al.} 2010; Lopes de Oliveira \rm{et al.}
2010 and our study), a significant fraction ($\geq$50\%) of systems
previously identified as FGs fail to meet the criteria. In fact,
recent works (Santos \rm{et al.} 2007; Zibetti \rm{et al.} 2009; our
study) have shown that even the Jones \rm{et al.} (2003) study, in
which the FG criteria were defined, underestimated the virial radius
of many of the systems it reports, and therefore \emph{also} used a
radius less then 0.5R$_{200}$. This is, most poignantly, found to be
true for the \emph{prototypical} system identified in Ponman \rm{et
  al.}  (1994) (i.e. RXJ1340), which fails a strict application of the
Jones et al. (2003) criteria.

We are therefore left with the choice of either discarding more than
50\% of the FG data in the literature (including the prototypical
system; RXJ1340) or allowing a relaxation of the selection
criteria. Such a relaxation amounts to simply accepting that the
systems under discussion represent the most extreme cases of large
magnitude gaps identified to date. Since, as noted in many of the
above papers, the original criteria are somewhat arbitrary, we have
elected to take the same approach as that in the majority of
literature studies and relax the FG criteria. We therefore simply note
that 8 out of 10 of the groups in our sample meet a definition of
fossil groups as systems with m$_{12}>$2.0~mag within 0.5R$_{200,S}$
(which corresponds to $\sim$0.33R$_{200,X}$), while the two systems
that fail these criteria still exhibit large gaps within this radius.\\

\begin{table*}
\begin{centering}
\caption{The richness (N$_{200}$; see Miller \rm{et al.} 2011), r band
  luminosity of the central galaxies (L$_{r,bcg}$) and the total r
  band luminosities within R$_{200,X}$ (L$_{r,tot}$) are given. The
  luminosity of the BCG is also given as fraction of the total
  optical light (f$_{bcg}$). The magnitude gaps between first and
  second ranked galaxies within R$_{200,S}$ -- i.e. those used in the
  selection process - are represented by m$_{12,S}$, while the
  magnitude gaps found within R$_{200,X}$ are represented by
  m$_{12,X}$. Finally, the r band dynamical mass-to-light ratios are
  presented.}
\begin{tabular}{|l|c|c|c|c|c|c|c|}  
\hline
Group    &N$_{200}$&log L$_{r,bcg}$&log L$_{r,tot}$&f$_{bcg}$  &m$_{12,S}$   & m$_{12,X}$   & log(M$_{dyn}$/L$_{r,tot}$) \\
         &        &(L$_{r,\odot})$ &(L$_{r,\odot}$)&          &(mag)        &  (mag)       & (M$_{\odot}$/L$_{r,\odot}$)\\
\hline                                                                                  
SDSS J0906    &  9    &  11.42        &  11.85        &0.38  &   3.09 & 3.09          &2.40 $\pm$ 0.22\\
SDSS J1045    & 13    &  11.44        &  12.14        &0.20  &   2.00 & 2.00          &2.44 $\pm$ 0.16\\
SDSS J1136    & 10    &  11.40        &  12.10        &0.21  &   2.25 & 0.58          &2.20 $\pm$ 0.26\\
SDSS J0856    & 16    &  11.28        &  12.05        &0.18  &   2.25 & 1.67          &2.21 $\pm$ 0.20\\
SDSS J1017    & 12    &  11.37        &  11.75        &0.41  &   2.72 & 1.88          &2.48 $\pm$ 0.29\\
RXJ  J1256$^*$&  8    &  11.15        &  11.79        &0.24  &   1.53 & 1.34          &2.71 $\pm$ 0.41\\
RXJ  J1331    &  6    &  10.94        &  11.41        &0.34  &   1.85 & 1.85          &2.33 $\pm$ 0.27\\
RXJ  J1340    &  8    &  11.36        &  11.81        &0.36  &   2.78 & 1.31          &2.40 $\pm$ 0.15\\
RXJ  J1416$^*$& 28    &  11.76        &  12.20        &0.35  &   2.55 & 2.21          &2.65 $\pm$ 0.18\\
RXJ  J1552$^*$& 19    &  11.50        &  12.10        &0.25  &   2.27 & 1.10          &2.66 $\pm$ 0.31\\
\hline
Average       & 13    &  11.37        &  11.91        &0.29  &   2.33 & 1.70          &2.45         \\
\hline
\end{tabular}
\label{photo}
\end{centering}
\\
$^*$ The r band photometry of the galaxies marked by an
  asterisk were estimated in the i band and converted to the r band
  using the values given in Section \ref{photanal}.
\end{table*}

\begin{figure}
\centerline{\psfig{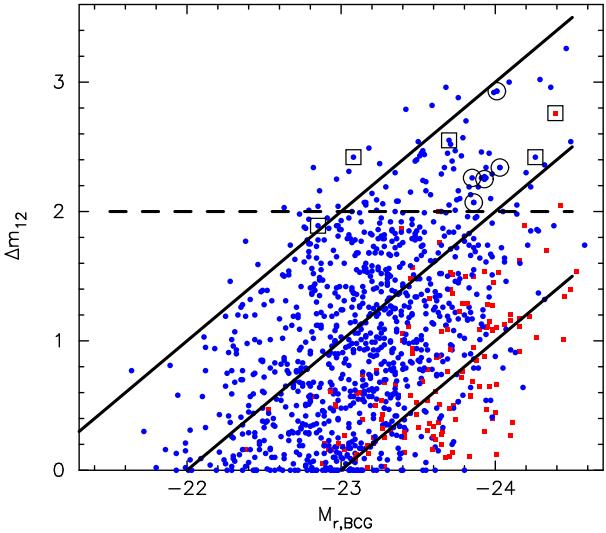}}
\caption{The luminosity gap (m$_{12,S}$) is plotted against BCG absolute
  magnitude. Lines of constant second ranked galaxy luminosity
  are marked by (diagonal) lines. These correspond to (from top to
  bottom) M$_2$=-21, -22 and -23~mag. Systems reported in this
  work are identified by circles (the SDSS sample) and squares (the
  RXJ sample). Points are coloured to indicate richness, with low
  richness groups (N$_{200}\le$25) in blue and high richness
  groups(N$_{200}>$25) in red.}
\label{select}
\end{figure}

Next, let us consider the m$_{12,S}$--L$_{bcg}$ data from the maxBCG
catalogue. These are shown in Fig. \ref{select} (N.B. These data are
also reported, in a slightly different form, in Miller \rm{et al.}
2011). In this figure, diagonal lines mark the loci of second ranked
galaxies of constant luminosity. Individual data points are coloured
by richness (blue for N$_{200}<$25, red for N$_{200}>$25; see Miller
\rm{et al.} 2011 for the definition of richness). The groups detailed
in this work are identified with circles (the SDSS sample) and squares
(the RXJ sample). There are a total of 1128 systems shown in this
plot.  It should be noted that these data were taken directly from the
maxBCG catalogue. The BCG luminosities are therefore not scaled to the
cosmology generally used in this paper.\\

Examination of Fig. \ref{select} reveals three important properties of
all systems exhibiting m$_{12}>$2~mag (i.e. not just those analysed in
this work):

i) Most of the BCGs in these systems are extremely bright. Indeed,
many are amongst the brightest in the whole sample, and very few
systems exhibit M$_{r,BCG}>$--23~mag.

ii) Most of these systems possess very low luminosity second ranked
galaxies, with very few systems exhibiting M$_{r,2}\le$--22~mag.

iii) Nearly all systems with m$_{12}>$2~mag exhibit low
richness. Indeed, the average richness of all 93 systems with
m$_{12}>$2~mag is only 13.0.  Only two systems with m$_{12}\ge$2~mag
exhibit N$_{200}>$25. These include J1416 (a member of our sample),
which has N$_{200}$=28, and another system with N$_{200}$=48.\\

It is interesting to note that the last of these results implies that
the low richness criterion applied in the selection of the SDSS sample
(Section \ref{ss}) was largely redundant. It should also be noted that
the high luminosities of the BCGs in our samples are highly selection
biased, as the SDSS sample was \emph{specifically} selected to contain
only systems with bright BCGs (see Section \ref{ss}), whereas the RXJ
sample was biased towards high luminosity BCGs by the selection of
systems with low ratios of X-ray to BCG optical luminosity (in systems
already known to be bright in the X-ray), as well as the selection of
``\emph{.......system[s] dominated by a single galaxy}'' (Jones et
al. 2003). However, this selection criterion too is largely redundant,
as the simple selection of samples with m$_{12}>$2~mag automatically
ensures a significant population of bright BCGs.

Now, the findings above -- that FGs are found in systems with bright BCGs
and low luminosity second ranked galaxies -- is perhaps not
surprising. However, the realisation that these high mass systems,
with appropriately bright BCGs, are low richness points to
a new interpretation of these objects

An important consideration in this new interpretation of fossil
systems is that there is a causal link between points ii and iii
above.  Namely, that low richness systems are \emph{expected}, on
average, to have low luminosity second rank galaxies, simply due to the
effect of sparse sampĺing of the Schechter function. This effect is
clearly demonstrated in the bottom panel of Fig. 10 of Hansen \rm{et al.}
(2005), in which the number of bright galaxies
present in systems declined rapidly as richness decreases from moderate to low
values. To further develop this point, in the next section we
present a completeness corrected, composite luminosity function for
the five systems in the SDSS sample (which, unlike the RXJ sample,
have high completeness out to the virial radius).

\subsection{The luminosity function}
\begin{figure}
\vspace{3mm} \centerline{\psfig{figure=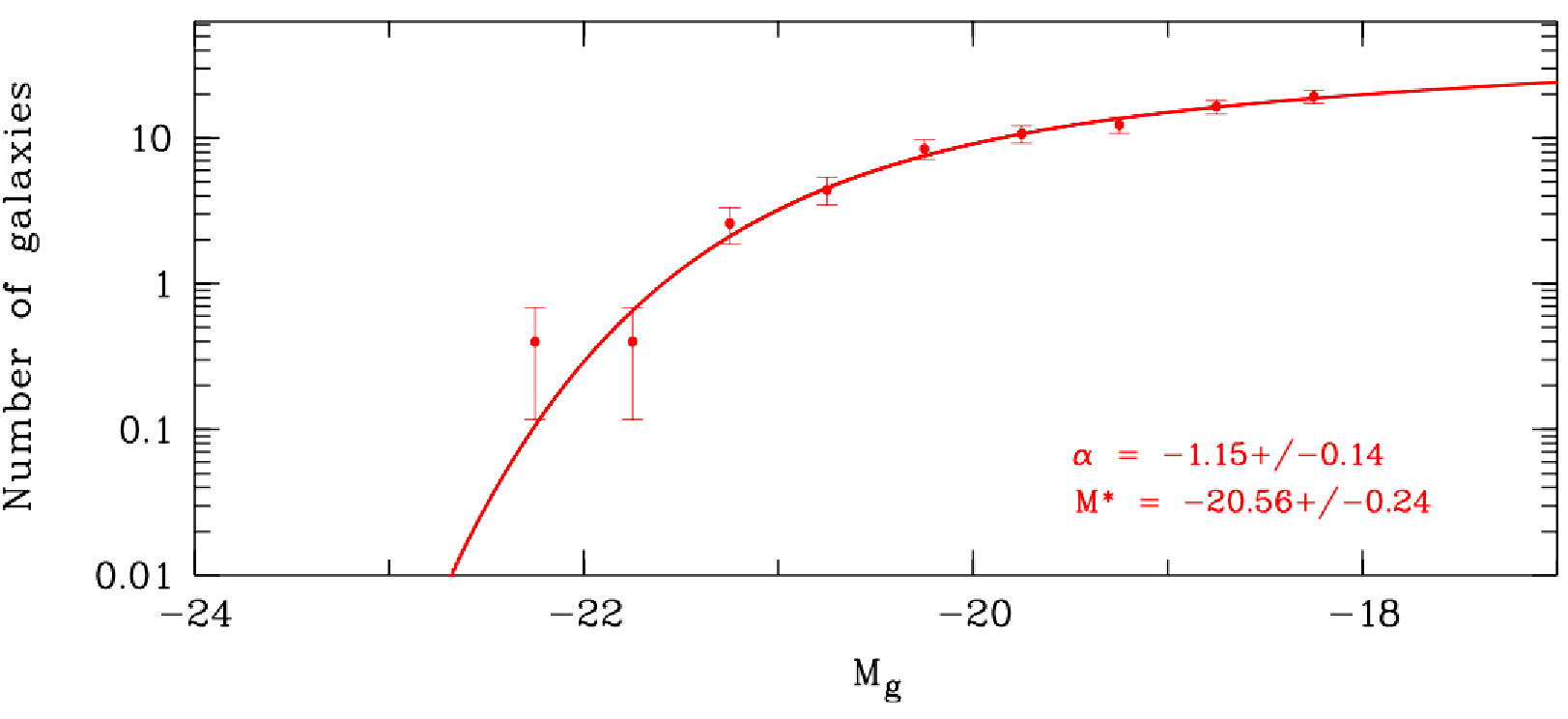,width=8cm,angle=0}}
\vspace{3mm}
\centerline{\psfig{figure=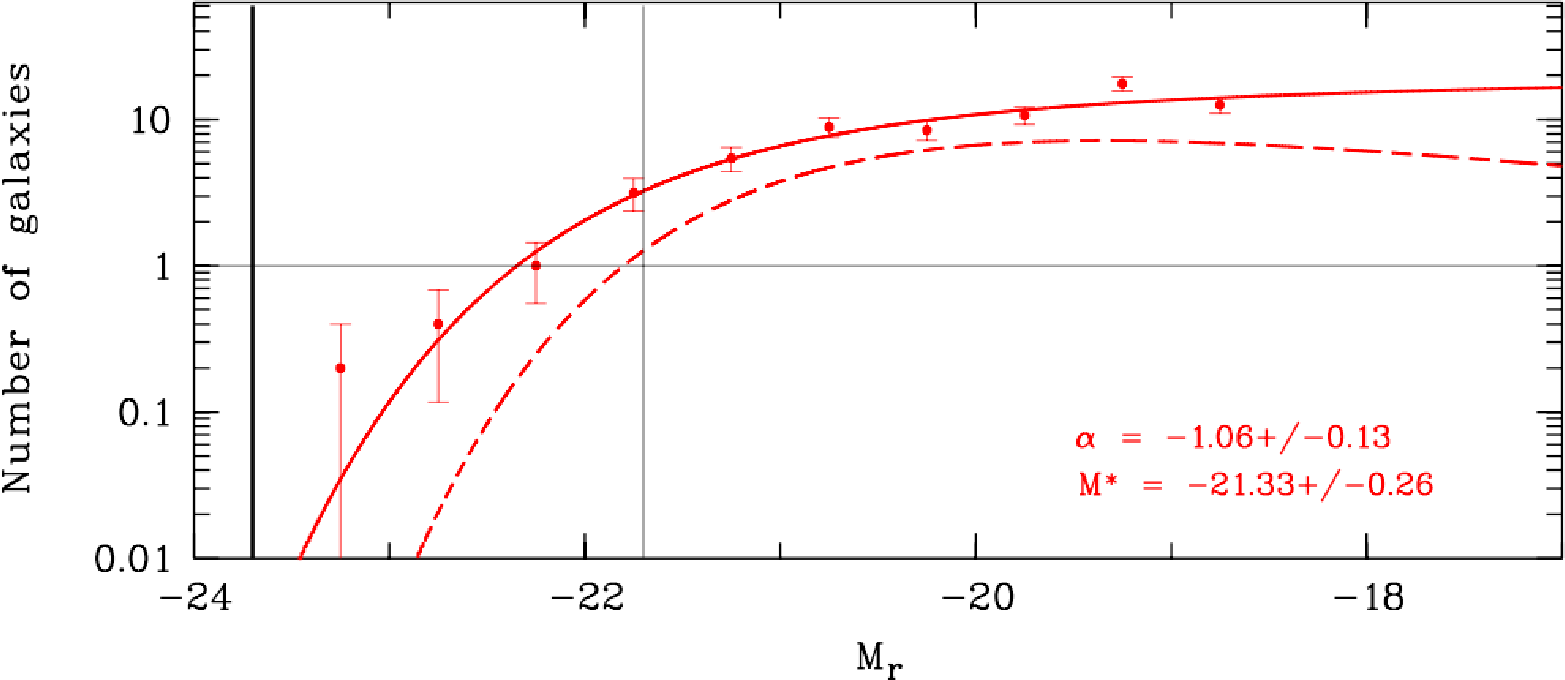,width=8cm,angle=0}}
\vspace{3mm}
\centerline{\psfig{figure=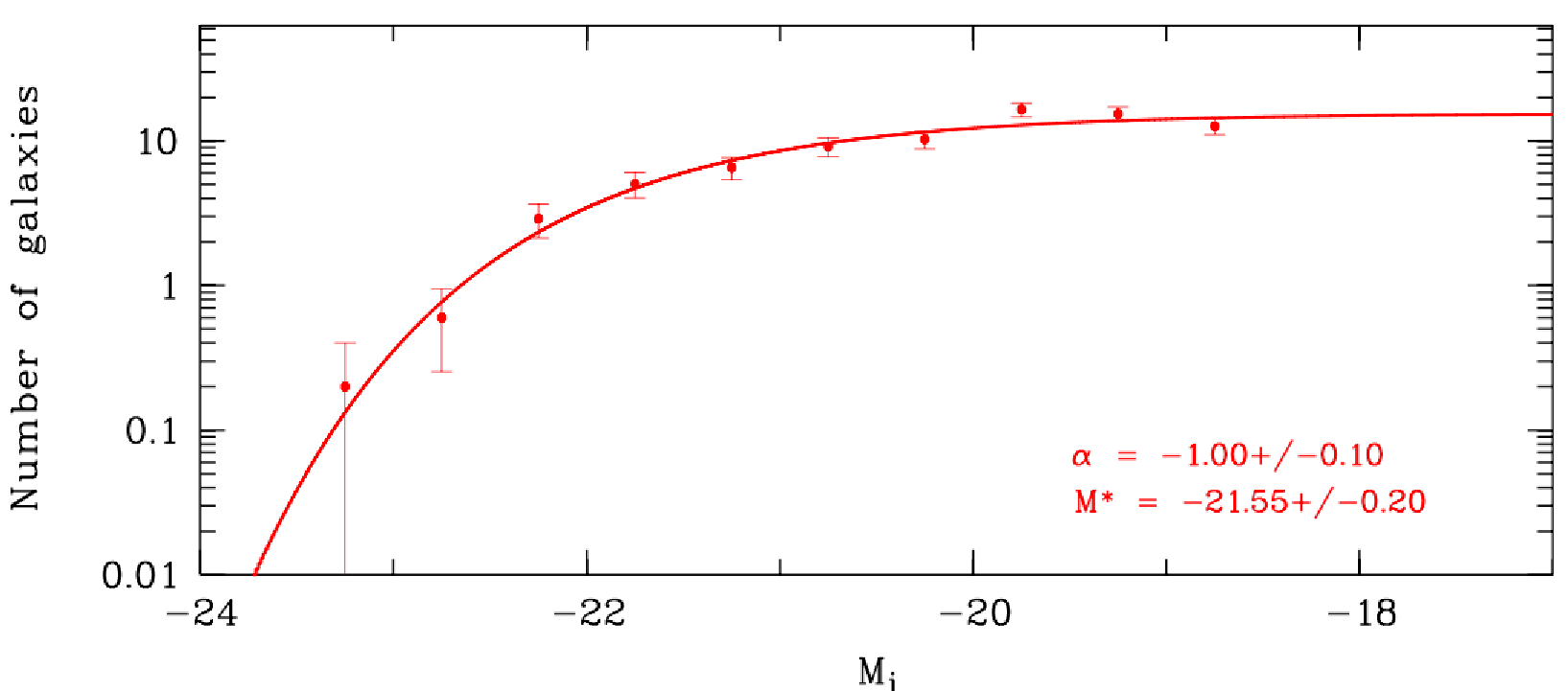,width=8cm,angle=0}}
\caption{Composite luminosity function for five fossil group
  candidates, J0906, J1045, J0856, J1136 and J1017. The solid red line
  is the composite (average) luminosity function within R$_{200,X}$
  for all five systems. The dashed line in the r band plot is the
  composite (average) luminosity function within 0.5R$_{200,X}$. The
  dark vertical line in the r band plot marks the average BCG luminosity,
  while the faint line marks the point two magnitudes fainter.}
\label{lf}
\end{figure}

The results of our analysis of the group luminosity functions (LFs) of
the SDSS sample in each of the g, r and i bands are shown in
Fig. \ref{lf}.

This analysis is limited by the completeness limit of the SDSS
photometry of about $\sim$ 21 mag in the $r$-band.  For the composite
LF, we assume the limit to the absolute magnitude to be -18 for the
g-band, and -18.5 for the r and i-band.  These conservative limits
were set to make sure that the individual LFs were considered inside
reasonable completeness limits, before including them in the composite
luminosity functions.  All galaxies meeting these limits and within
R$_{200,X}$ of each system were included, and the LF averaged. The
number of galaxies (y-axis of Fig. \ref{lf}) therefore represents the
number of galaxies per magnitude found within R$_{200,X}$ of a single,
average group.

The best fit for $\alpha$ and M$^*$ for the three bands are -1.15 and
-20.56~mag for g, -1.06 and -21.33~mag for r and -1.00 and -21.55~mag
for i. The derived LFs are shown in Fig. \ref{lf} as solid red lines.
Also shown in the r band plot is the LF within 0.5R$_{200,X}$) which
exhibits $\alpha$ and M$^*$ of -0.69 and -20.62. The error bars on our
derived LFs are large, due to the relatively small number of galaxies
used in their construction. The results are therefore consistent with
a broad range of literature studies. We do, however, note a
particularly good agreement between the $\alpha$ and M$^*$ values that
we derive in the r band within R$_{200,X}$ and the values of Blanton et al
(2003) for $\sim$150,000 galaxies in the SDSS spectroscopic survey.

It should be remembered that the LFs presented in Fig. \ref{lf} are the
\emph{average} of five individual groups. The y-axis of Fig. \ref{lf}
therefore represents the expectation for the number of galaxies per
magnitude bin of a single, average group.

Now, we are concerned with describing the m$_{12,S}$ gaps which are
defined to be within 0.5R$_{200}$. Our analysis therefore proceeds by
considering the r band LF within 0.5R$_{200,X}$ (Fig. \ref{lf}, middle
plot, dashed line). It is evident from this plot that the expectation
of the number of galaxies per magnitude is below 1.0 \emph{over the
  entire 2 magnitude range immediately below the luminosity of the
  BCG}. This clearly indicates that m$_{12}$ gaps are likely to be
large in such low richness, bright BCG systems. The effect can be
quantified by integrating along the LF over this 2~mag range. The
derived value of 0.4 indicates that $\sim$60\% of all such systems
will posses \emph{no} galaxies within this magnitude range (and
therefore possessing m$_{121,S}>$2.0~mag). Analysis of the data shown
in Fig. \ref{select} shows that, in fact 50\% of the 62 low richness
(N$_{200}<$25) systems with bright BCGs (M$_{r,bcg}<-23.5$~mag) possess
m$_{121,S}>$2.0~mag, in reasonably good agreement with our estimate from
the r band LF of 60\%.

This result is in sharp contrast with Jones \rm{et al.} (2003) who
performed Monte Carlo simulations using the LF of MKW/AWM clusters
from Yagamata \& Maehara (1986), finding an extremely low incidence of
systems with m$_{12}>$2.0~mag. However, it is not clear from their
paper what value of BCG luminosity was used. We therefore performed
our analysis again, this time using the Yagamata \& Maehara values for
M$^*$, $\alpha$ and BCG luminosity (-21.57~mag, -1.07 and -23.0~mag,
respectively).  Our analysis indicates that using these values the
expectation of the number of galaxies within 2 magnitudes of the BCG
is 3.4. We therefore confirm that such an analysis results in
extremely low probabilities of finding fossil groups. Indeed, assuming
simple Poisson statistics, the average of 3.4 galaxies within 2
magnitudes of the BCG would suggest fossil systems with
m$_{12}>$2.0~mag to be 2$\sigma$ events, consistent with the low
numbers reported in Jones et al. (2003).

The cause of the disparity between the Jones \rm{et al.} results and
ours is therefore likely to be due to the difference in the gaps
between M$^*$ and L$_{bcg}$ in the two studies, i.e. 2.5~mag in
our study, but only 1.5~mag in the Yagamata \& Maehara (1986). We
therefore conclude it to be likely that the Jones \rm{et al.} (2003)
analysis did not consider the extremely bright BCGs found in fossil
systems.

In summary, our analysis shows that large m$_{12}$ gaps are an
expected feature in low richness systems that host bright BCGs, and
that this effect alone can account for the properties of the fossil
systems in our study, without recourse to additional processes such as
dynamical friction.

\subsection{Synthesis}
In this section we draw together the various strands of our analysis
in order to gain a clearer insight into the nature of the fossil
systems that we have investigated, and to identify outstanding issues.

First, let us recall the two most important conclusions of our
analysis of the scaling relations of fossil system. Namely, that these
systems possess high masses and, despite the high luminosities of
their BCGs, low \emph{total} optical luminosities (Section
\ref{scale}). 

Considering the bright BCG luminosities and high system masses, it can
be seen from the plot of BCG luminosity against mass (M$_{200,dyn}$)
in Hansen \rm{et al.} (2009; their Fig. 13) that the values found for
our fossil groups (average L$_{bcg}$=2.3$\times$10$^{11}$L$_{\odot}$,
M$_{200}$=2.3$\times$10$^{14}$M$_{\odot}$) are consistent with values
found for normal systems in the SDSS. Indeed, Fig. 13 of Hansen \rm{et
  al.} (2009) indicates that, for a mass of
2.3$\times$10$^{14}$M$_{\odot}$, a typical BCG luminosity is
2$\times$10$^{11}$L$_{\odot}$ (after the Hansen \rm{et al.} data is
k-corrected to z=0 and adjusted to the cosmology used in this
work). In other words, the BCG luminosities in these fossil systems
are entirely consistent with their masses, but are inconsistent with
either their richnesses or total optical luminosities.

The difference between fossils and normal systems can also be seen by
examining the plot of the fraction of optical light in the BCG
(f$_{bcg}$ in our Table \ref{photo}) to M$_{200}$ of SDSS groups and
clusters as shown in the top panel of Fig. 14 of Hansen \rm{et al.}
(2009). For masses appropriate for our sample, this plot shows that
normal systems with masses appropriate to our fossil sample
(2.3$\times$10$^{14}$M$_{\odot}$) possess f$_{bcg}\approx$0.1 (again
after the Hansen \rm{et al.} data is adjusted to the cosmology used in this
work). Comparison of this value to the value of $\sim$0.3 found in fossil
systems therefore again suggests a factor $\sim$3 under-luminosity in
fossils systems compared to normal systems, consistent with the value
found by consideration of the L$_r$--$\sigma$ plot (Fig. \ref{sr} and
Section \ref{scale}).

Now, given that we find the luminosities of the BCGs to be comparable
between fossil and normal samples of the same mass, the discrepancy in
the optical luminosity must be due to a significant under-abundance of
\emph{non}-BCG galaxies. Indeed, simple arithmetic shows that, if the
whole deficit is due to the lack of non-BCG galaxies, then, in order
that the \emph{total} luminosity be $\sim$3 times lower than normal
systems, fossil must contain only $\sim$25\% of the non-BCG galaxies
found in a normal cluster of the same mass.\\

\subsection{The role of dynamical friction}
Given our results it is interesting to look at the possible role of
dynamical friction in generating the large m$_{12}$ gaps in our
fossil sample. Our considerations are based on the luminosity function
derived for the average of the five groups in the SDSS sample, as
detailed above.

Consider the r band LFs in Fig. \ref{lf}, if we assume
that dynamical friction has caused even a single bright galaxy to be
``cannibalised'' by the central BCG of each group, then replacing this
galaxy in the LF within R$_{200}$ (i.e considering what this LF looked
like \emph{before} dynamical friction has had its effect) results in a
LF that shows a significant excess of bright galaxies with respect to
any reasonable Schechter function, i.e. there would be as many, or
\emph{more}, galaxies with luminosity $\sim$--23 ~mag than there
galaxies with luminosity $\sim$--22 mag.  The problem is even more
pronounced if the putative cannibalised galaxies are replaced in the LF
within 0.5R$_{200}$ (dashed line Fig. \ref{lf}; middle panel).

Therefore, while our data can not speak to the role of dynamical
friction in the building of the LF and BCG at early times, the data
does suggest that, if dynamical friction \emph{has} played a part in
generating the large m$_{12}$ gaps and bright BCGs at later times,
then our sample of fossil groups must have started with abnormal LFs.

\section{Discussion}
The picture painted by our analysis can then be summarised as follows:
fossil groups differ significantly from non-fossils systems of the
same mass \emph{only} in that they exhibit a large under-abundance of
non-BCG galaxies. We note that this description is highly efficient in
that it simultaneously describes the similarities and differences
between fossil and normal systems for a host of observables
(e.g. N$_{200}$, m$_{12}$, L$_{bcg}$, L$_{tot}$, L$_X$, T$_X$ and
$\sigma$)

In the light these conclusions, a number questions (but, unfortunately,
not many answers) immediately present themselves;

- \emph{Where are all the missing bright baryons?}\\ 
There are three immediately apparent ways to account for the
``missing'' baryons:

i) They have been expelled from the system (although it is difficult
to see how this could be accomplished without a significant loss of X-ray
emitting gas, which has not been detected).

ii) They are ``hidden'' -- possibly locked up in the hot X-ray gas or
the warm/hot intergalactic medium, suggesting a low galaxy
formation efficiency. Alternatively, they could be ``hidden'' from our
luminosity budget as intra-cluster light (although this seem unlikely,
as the high velocity dispersions and low galaxy number densities
exhibited by the groups in our sample suggest the interactions that
generate intra-cluster light would be rare and weak).

iii) They were never present at all, with the systems forming in
regions of space deficient in baryons (although it is difficult to see
how the bright BCGs could have formed in such circumstances).

- \emph{Are fossils really fossils? I.e. are they truly \emph{old}?}\\ 
It is difficult to see how such low mass-to-light systems could have
formed \emph{recently}. Significant merger/accretion activity would
also seem to be ruled out as this would have both ameliorated the high
mass-to-light ratios and provided a significant source on non-BCG
galaxies (the one thing above all else that these systems lack). It
therefore seems safe to conclude that these are indeed ancient
systems, and that they are indeed worthy of the title \emph{fossil}
groups.

- \emph{What do our results mean for studies that utilise cosmological 
N-body/semi-analytic modeling to address issues surrounding fossil groups?}\\
As far as the authors are aware, no such study to-date has identified
fossils as being associated with low richness systems. Whether this is
a failure of the studies themselves or rather represents a failure in
the baryonic physics in the semi-analytical models used in the
cosmological simulations remains to be seen.

- \emph{By what mechanism could the BCGs in the low richness systems
  of fossil groups achieve the same mass as those in much richer
  systems?}\\ 
Our results appear to present a challenge to the currently accepted
paradigm of BCG formation through hierarchical clustering within the
host halo (e.g. de Lucia \& Blaizot 2010), since the massive
(luminous) BCGs found in our sample of fossil systems appear to have
formed in extremely low galaxy number-density environments, and should
therefore have been relatively starved of the raw material necessary
for such a hierarchical assembly path.\\

All of these issues clearly need addressing in the near future if we
are to establish a coherent picture of how the formation of fossil
systems differs from normal systems.

\section{Conclusions}
\label{concs}
We present a kinematic analysis of ten fossil group candidates, five
of which have been previously identified as fossil groups in the
literature. The other five candidates investigated were optically
selected from the maxBCG catalogue of Koester \rm{et al.} (2007),
spectroscopically observed with the Magellan IMACS instrument and
followed up with Chandra X-ray snapshot observations.  For these 10
groups, between 10 and 64 galaxies (with an average of $\sim$33) are
confirmed as group members within R$_{200,X}$. This study therefore
represents the deepest study of a significant number of fossil systems
to-date.

We confirm previous findings that the majority of the FGs identified
to-date lie in the regions of X-ray luminosity, X-ray temperature and
velocity dispersion scaling relations occupied by galaxy
\emph{clusters} rather than groups. Since all three of these
parameters (L$_X$, T$_X$ and $\sigma$) can be used as proxies for
mass, and all three yield masses consistent with cluster masses
($\sim$10$^{14}$M$_{\odot}$, or greater), we can be confident in these mass
estimates. We find that the luminosities of the brightest cluster
galaxies in our sample are also consistent with these high masses,
lending further support to this finding.

However, there is one parameter that is \emph{not} consistent with
cluster values -- namely the total optical luminosities of these
systems. We find that the fossil groups in our sample are, on average,
under-luminous by a factor $\sim$3 with respect to galaxy clusters of
the same mass. High mass-to-light ratios have been noted in previous
works (e.g. Jones \rm{et al.} 2003; Yoshioka \rm{et al.} 2004;
Cypriano \rm{et al.} 2006; Mendes de Oliveira \rm{et al.} 2006;
KPJ07), but no firm conclusions were drawn from these relatively small
samples. Here, however, we find this to be essentially the defining
feature of fossil systems, showing that these systems are
characterised by their possession of only $\sim$25\% of the non-BCG
galaxies found in normal systems of the same mass. We show that this
low richness can simultaneously account for the large m$_{12}$ gaps
\emph{and} the high mass-to-light ratios.

We note that the \emph{none} of the paradigms for the formation of
fossils (and particularly the paradigm of cannibalism of bright
central galaxies by the BCG) predict such high masses coupled with low
luminosities. Our findings therefore suggest that a new paradigm for
the formation and evolution of fossil groups is required.

\newpage
\noindent{\bf References}\\
Aguerri J.A.L., et al., 2011, A\&A, 527, 143\\
Bauer A.E., Gr\"{u}tzbauch R., J{\o}rgensen I., Varela J., Bergmann M., 2011, MNRAS, 411, 2009\\
Becker M.R., et al., 2007, ApJ, 669, 905\\
Beers T.C., Flynn K., Gebhardt K., 1990, AJ, 100, 32\\
Bertin E., Arnouts S., 1996, A\&AS, 117, 393\\
Blanton et al., 2003, AJ, 592, 819\\
Carlberg R.G., et al., 1997, ApJ, 485, L13\\
Cypriano E.S., Mendes de Oliveira C.L., Sodr\'{e} L., Jr. 2006, AJ, 132, 514\\
Dariush A.A., Khosroshahi H.G., Ponman T.J., Pearce F., Raychaudhury S., Hartley W., 2007, MNRAS, 382, 433\\
Dariush A.A., Raychaudhury S., Ponman T.J., Khosroshahi H.G., Benson A.J., Bower R.G., Pearce F., 2010, MNRAS, 405, 1873\\
de Lucia G., Blaizot J., 2007, MNRAS, 375, 2\\
D\'{e}mocl\`{e}s J., Pratt G.W., Pierini D., Arnaud M., Zibetti S., D'Onghia E., 2010, A\&A, 517, 52\\
D'Onghia E., Sommer-Larsen J., Romeo A.D., Burkert A., Pedersen K., Portinari L., Rasmussen J., 2005, ApJ, 630, L109\\
Fujita Y., 2004, PASJ, 56, 29\\
Girardi M., Giuricin G., Mardirossian F., Mezzetti M., Boschin W., 1998, ApJ, 505, 74\\
Girardi M., Manzato P., Mezzetti M., Giuricin G., Limboz F., 2002, ApJ, 569, 720\\
Gunn J.E., Gott J.R., 1972, ApJ, 176, 1\\ 
Hansen S.M., McKay T.A., Wechsler R.H., Annis J., Sheldon E.S., Kimball A., 2005, ApJ, 633, 122\\
Hansen S.M., Sheldon E.S., Wechsler R.H., Koester B.P., 2009, ApJ, 699, 1333\\
Helsdon S.F., Ponman T.J., 2003, MNRAS, 340, 485\\
Hilton M., et al., 2010, ApJ, 718, 133\\
Johnston et al., 2007, preprint (arXiv:0709:1159\\
Jones L.R., Ponman T.J., Horton A., Babul A., Ebeling H., Burke D.J., 2003, MNRAS, 343, 627\\
Kawata D., Mulchaey J. S., 2008, ApJ, 672, L103\\
Khosroshahi H.G., Ponman T.J., Jones L.R., 2007, MNRAS, 377, 595 ({\bf KPJ07})\\
Koester et al., 2007, ApJ, 660, 239\\
Landolt A.U., 1992, AJ, 104, 340\\
Lopes de Oliveira R., Carrasco E.R., Mendes de Oliveira C., Bortoletto D.R., Cypriano E., Sodré L. Jr., Lima Neto G.B., 2010, AJ, 139, 216\\
Mendes de Oliveira C.L., Cypriano E.S, Sodr\'{e}, L., Jr. 2006, AJ, 131, 158\\
Mendes de Oliveira C.L., Cypriano E.S., Dupke R.A., Sodr\'{e} L., Jr. 2009, AJ, 138, 502\\
Miller et al., 2011, preprint (arXiv:/XXXX.XXXX)\\
Maurogordato S., Sauvageot J.L., Bourdin H., Cappi A., Benoist C., Ferrari C., Mars G., Houairi K., 2010, A\&A, submitted (astro-ph:1009.1967)\\
Navarro J.F., Frenk C.S., White S.D., 1995, MNRAS, 275, 720\\
Osmond J.P.F., Ponman T.J., 2004, MNRAS, 350, 1511\\
Ponman T.J., Allan D.J., Jones L.R., Merrifield M., McHardy I.M., Lehto H.J., Luppino G.A., 1994, Nature, 369, 462\\
Popesso P., Biviano A., B\"{o}hringer H., Romaniello M., 2007, A\&A, 464, 451\\
Ramella M., Boschin W., Geller M.J., Mahdavi A., Rines K.,  2004, AJ, 128, 2022\\
Rasmussen J., Ponman T.J., Mulchaey J.S., 2006, MNRAS, 370, 453\\
Rozo E., et al., 2009, ApJ, 703, 601\\
Rykoff E.S., et al., 2008, MNRAS, 387 L28\\
Santos W.A., Mendes de Oliveira C., Sodr\'{e} L., Jr. 2007, AJ, 134, 1551\\
Schechter P., 1976, ApJ, 203, 297\\
Sheldon E.S., et al., 2009, 703, 2217\\
Smith G.P., et al., 2010, MNRAS, 409, 169\\
Strazzullo V., et al., 2011, A\&A, 524, 17\\
Vikhlinin A., McNamara B.R., Hornstrup A., Quintana H., Forman W., Jones C., Way M.,  1999. ApJ. 520, L1\\
Voevodkin A., Borozdin K., Heitmann K., Habib S., Vikhlinin A., Mescheryakov A., Hornstrup A., Burenin R., 2010, ApJ, 708, 1376\\
Wu X.-P., Xue Y.-J., Fang L.-Z., 1999, ApJ, 524, 22\\
Yagamata T., Maehara H., 1986, PASJ, 38, 661\\
Yoshioka T., Furuzawa A., Takahashi S., Tawara Y., Sato S., Yamashita K., Kumai Y., 2004,  2004, Advances in Space Research, 34, 2525\\
Zhang Y.-Y., Andernach H., Caretta C.A., Reiprich T.H., B\"{o}hringer H., Puchwein E., Sijacki D., Girardi M., 2011, A\&A, preprint (arXiv:1011.3018)\\
Zibetti S., Pierini D., Pratt G.W., 2009, MNRAS, 392, 525\\

\noindent

\noindent{\bf Acknowledgments}\\ 

This work is based on observations made with the 6.5 m Magellan/Baade
telescope, a collaboration between the Observatories of the Carnegie
Institution of Washington, University of Arizona, Harvard University,
University of Michigan, and Massachusetts Institute of Technology, and
at Cerro Tololo Inter-American Observatory, a division of the National
Optical Astronomy Observatories, which is operated by the Association
of Universities for Research in Astronomy, Inc. under cooperative
agreement with the National Science Foundation. The COSMOS pipeline
supplied by the Magellan consortium was used for data reductions. This
research also made use of NASA's Astrophysics Data System, as well as
IRAF and STARLINK software.  IRAF is distributed by the National
Optical Astronomy Observatories, which is operated by the Association
of Universities for Research in Astronomy, Inc. (AURA) under
cooperative agreement with the National Science Foundation.  Also
presented are observations obtained at the Gemini Observatory, which
is operated by the Association of Universities for Research in
Astronomy, Inc., under a cooperative agreement with the NSF on behalf
of the Gemini partnership: the National Science Foundation (United
States), the Science and Technology Facilities Council (United
Kingdom), the National Research Council (Canada), CONICYT (Chile), the
Australian Research Council (Australia), Minist\'{e}rio da Ci\^{e}ncia
e Tecnologia (Brazil) and Ministerio de Ciencia, Tecnolog\'{i}a e
Innovaci\'{o}n Productiva (Argentina). R.A.D acknowledges support from
NASA Grant NNH10CD19C and partial support from Chandra Award
No. GO9-0142A.  R.L.O acknowledges financial support from the
Brazilian agency FAPESP (Funda\c{c}\~{a}o de Amparo \`{a} Pesquisa do
Estado do S\~{a}o Paulo) through a Young Investigator Program (numbers
2009/06295-7 and 2010/08341-3).  R.N.P. also acknowledges financial
support from the Brazilian agency FAPESP (program number
2008/57331-0). E.S.C. also acknowledges FAPESP (program number
2009/07154-8-0) and CNPq.

\begin{appendix}
\section{The effects of substructure}
\label{app}

In this appendix we detail our analysis of the uncertainty in our
derived velocity dispersions in the three systems in our study that
possess the strongest indications of the presence of substructure
(J1017, J1256 and J1416, as described in Section \ref{rvs}).

\begin{figure}
\centerline{\psfig{figure=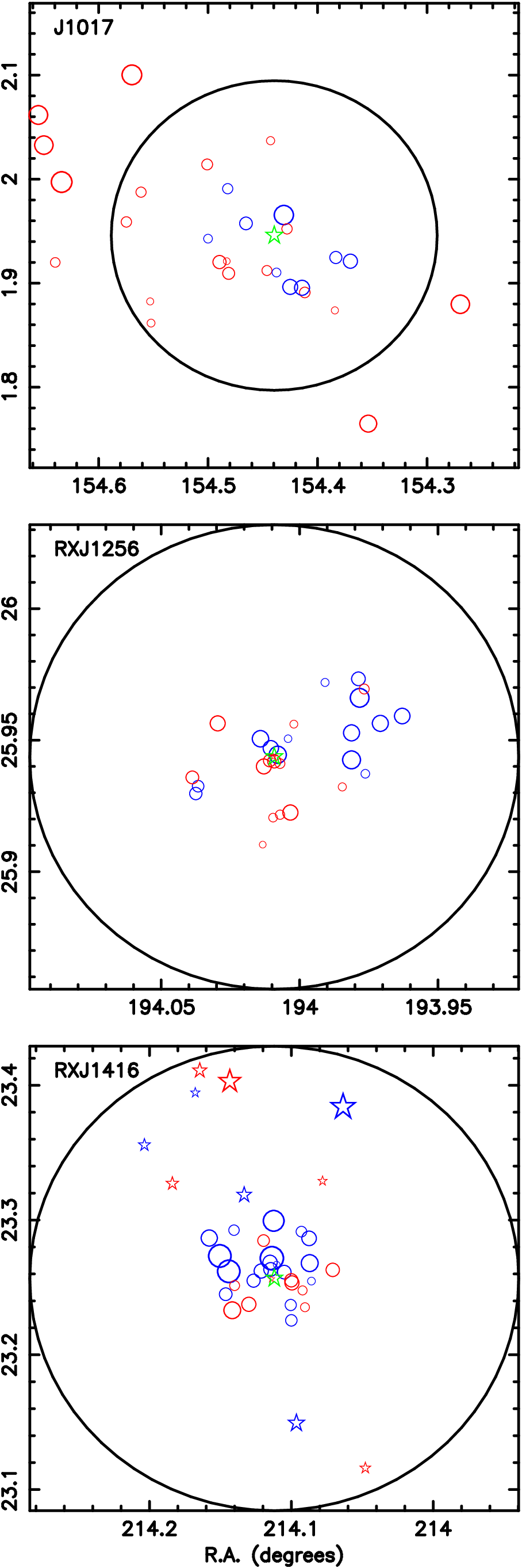,width=6.5cm,angle=0}}
\caption{The spatial distributions of galaxies with recession
  velocities within 2000~km~s$^{-1}$ of the BGG (green star) are
  shown. The symbol colour indicates the \emph {sense} of the
  recession velocity with respect to the BGG (red for redshift, blue
  for blueshift). Symbol size denotes the \emph{magnitude} of the
  recession velocity with large symbols representing greater absolute
  velocities. Stars identify
  galaxies whose recession velocities were taken from the SDSS
  spectroscopic catalogue. Solid circles mark the R$_{200,X}$ radius.}
\label{space1}
\end{figure}

Fig. \ref{space1} shows the spatial distribution of member galaxies of
these three groups. Circles representing R$_{200,X}$ are shown. Unfortunately,
even with the increase in the numbers of spectroscopically confirmed
members compared to previous studies, the incomplete spatial coverage
of our data results in it still being insufficient for a full
``friends-of-friends'' type of analysis. Here, we therefore
simply test the likely impact of potential substructures on our
derived parameters in the three systems including or excluding them
from our analysis.

\subsubsection{J1017}
In the case of J1017, there are a significant number of galaxies with
high recession velocities ($>$900~km~s$^{-1}$) located just outside
R$_{200,X}$. Examination of Fig. \ref{space1} shows them to be
located to the north-east and south and south-west of the group in a
configuration which suggests that these galaxies are unlikely to be
group members under the assumption that the group is
virialised. However, to test for the impact of including these
galaxies in our kinematic measurements, we re-calculated the velocity
dispersion and $\Delta$RV$_{BCG}$ including these galaxies. The effect
was to increase the measured velocity dispersion from 474$\pm$71 to
643$\pm$88~km~s$^{-1}$ (an increase in log $\sigma$ of
0.13~dex). Comparison of the dynamical R$_{200}$ (as defined in
the previous section and given in Table \ref{data}) with the X-ray
derived value (Table \ref{xray_data}) shows almost perfect
agreement. Therefore, the dynamical R$_{200}$ that would be
derived from the increased value of velocity dispersion when the
outlying galaxies are included would exceed the X-ray derived value by
nearly 40\%. Their inclusion also results in a significant decrease in
$\Delta$RV$_{BCG}$, with the value going from
+73$\pm$101~km~s$^{-1}$ to --129$\pm$125~km~s$^{-1}$.  Therefore,
while we can draw no firm conclusions, the data suggest that these
high radius, high recession velocity galaxies are not part of the
virialised system.

\begin{figure}
\centerline{\psfig{figure=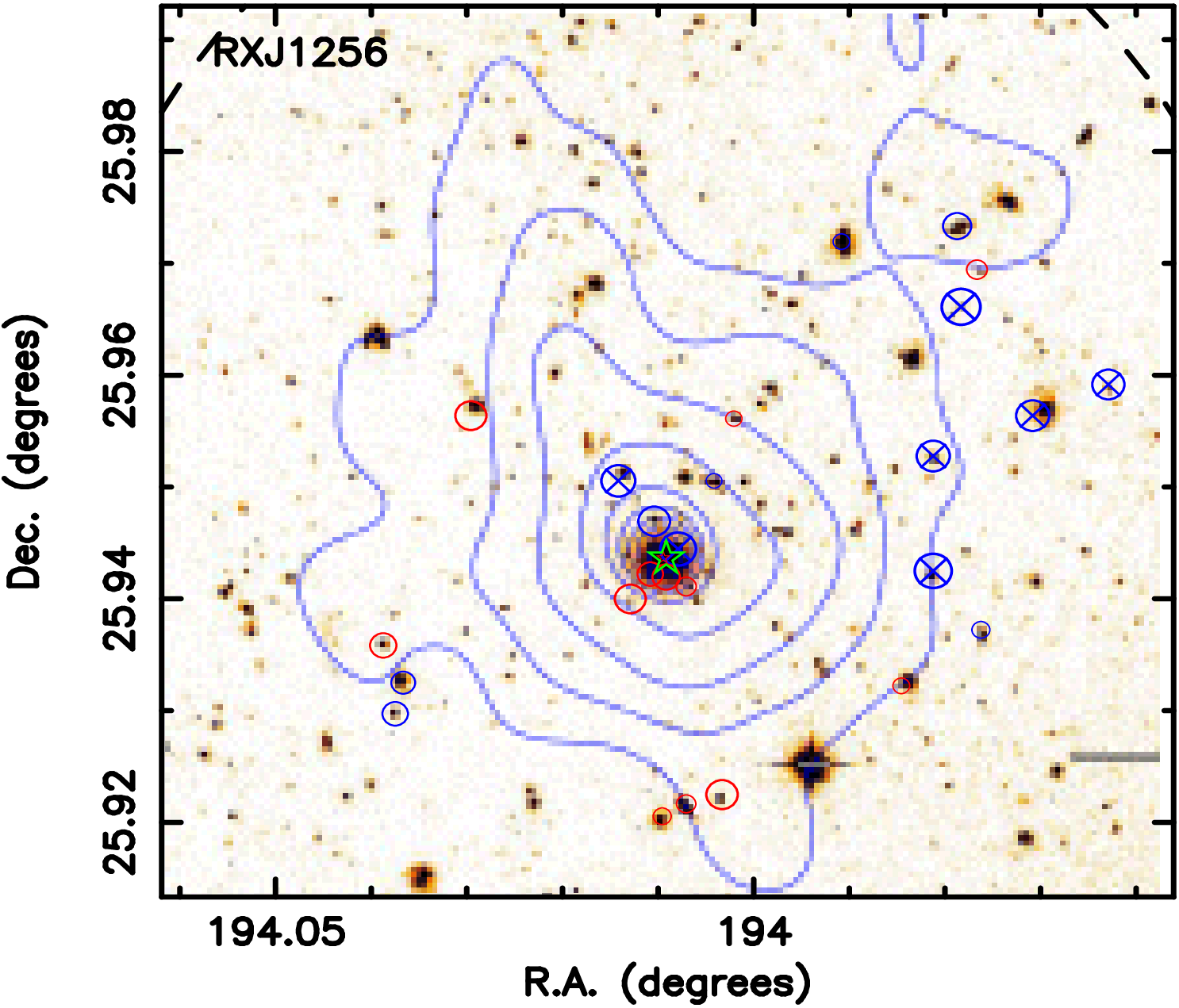,width=7cm,angle=0}}
\centerline{\psfig{figure=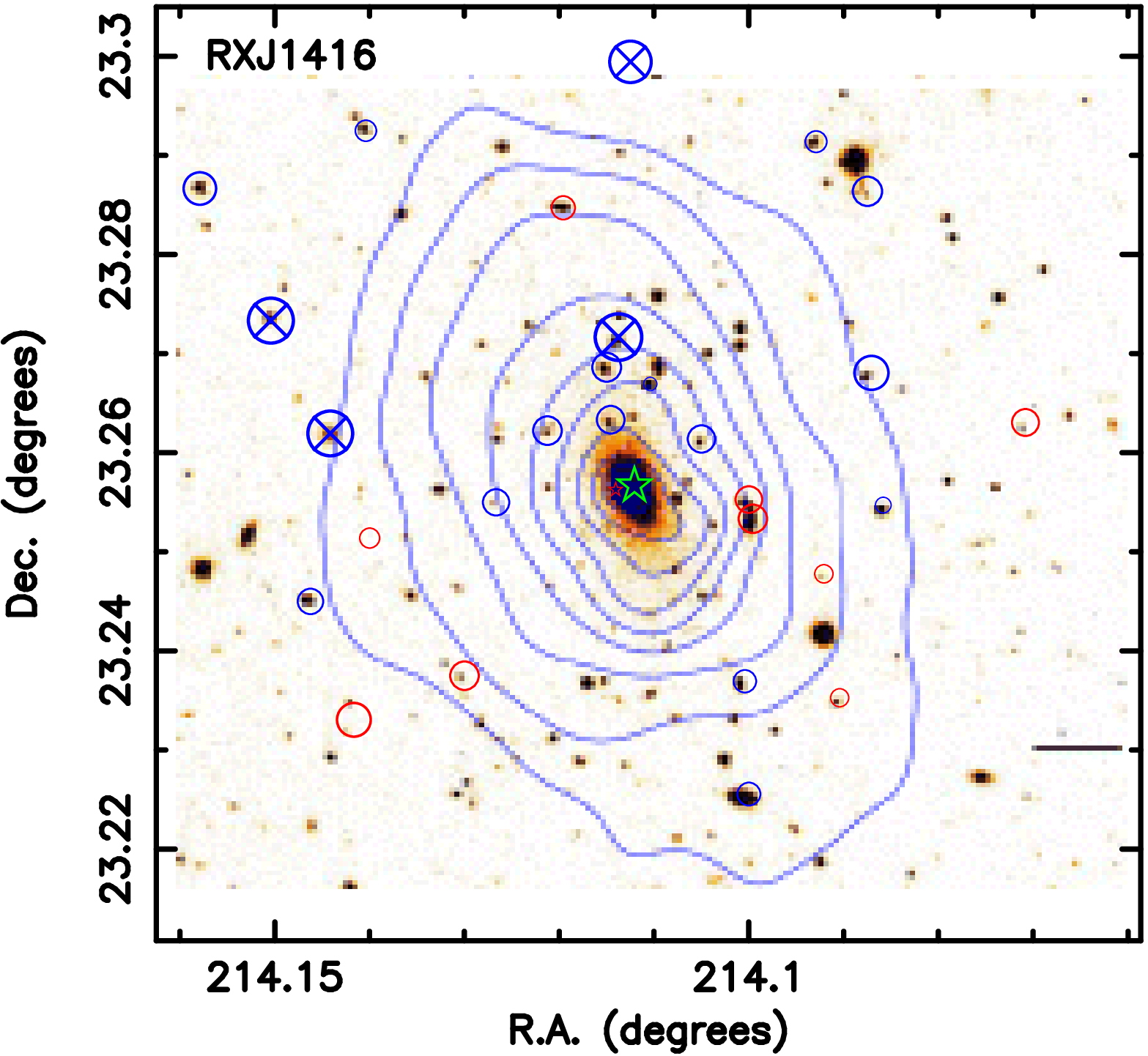,width=7cm,angle=0}}
\caption{Our recession velocity data for J1256 and J1416 are overlayed
  on Chandra X-ray contours from KPJ07.}
\label{xray_overlay}
\end{figure}

\subsubsection{J1256}
For J1256 we see an apparent excess of low recession velocity galaxies
(Fig. \ref{rvhistos}). Examination of Fig. \ref{space1} shows that
these galaxies all lie on one side of the group. Indeed, the group is
remarkably reminiscent of the ``bimodal'' clusters reported in
Maurogordato et al. (2010). The BCG also exhibits a large positive
recession velocity with respect to the group average of ($\Delta$RV$_{BCG}$; Table
\ref{data}). We therefore again re-calculated the velocity dispersion
and dynamical R$_{200}$, this time excluding galaxies with absolute
recession velocities less than --800~km~s$^{-1}$. This resulted in a
velocity dispersion of 449$\pm$70~km~s$^{-1}$ and a $\Delta$RV$_{BCG}$ of
--115$\pm$100~km~s$^{-1}$. The reduction in velocity dispersion of 28\%
(0.14 dex in log $\sigma$) results in a reduction in the dynamical
R$_{200}$ such that it becomes 20\% lower than the X-ray derived
value rather than the 16\% over-estimate of the value given in
Table \ref{data}. We are therefore unable to definitively state
whether these objects are part of the virialised structure on the
basis of the velocity dispersion measurements. In addition, while the
large positive $\Delta$RV$_{BCG}$ is eliminated, its replacement by the
relatively large negative value means that no firm
conclusions can be drawn from this parameter either. However,
examination of the X-ray contours (Fig. \ref{xray_overlay}) of this
group indicates no obvious sign of substructure associated with the
overall projected location of the highly blueshifted
galaxies. In fact, the X-ray isophotes seem to show an elongated
substructure towards the N-NE, consistent with some dynamical
turmoil, so that the sub-system of velocity outliers may be
the results of the previous interaction with another group and the
system may not be fully virialised. A more complete study of this
group is necessary to confirm these findings.

We also note that a close examination of the central galaxy of this
system reveals two extremely nearby (on the plane of the sky), relatively bright galaxies
that have not to-date been examined spectroscopically (either by us or
the SDSS), and whose membership therefore remains untested. The
brighter of these two galaxies is only 1.83~mag fainter in the r band
than the central galaxy.

\subsubsection{J1416}
In J1416 we see a substantial collection of low recession velocity
galaxies in Fig. \ref{rvhistos} which are separated in redshift space
from the remainder of the group by a significant gap.  Examination of
Fig. \ref{space1}, once again, shows all the low-z galaxies to lie on
one side of the group, again reminiscent of the ``bimodal'' clusters
reported in Maurogordato et al. (2010). The group also exhibits the
largest $\Delta$RV$_{BCG}$ in our sample.  We therefore
re-calculated the velocity dispersion and dynamical R$_{200}$, this time
excluding galaxies with absolute recession velocities less than
--1400~km~s$^{-1}$. Once again, we find the large $\Delta$RV$_{BCG}$ to be
eliminated (becoming +74$\pm$98~km~s$^{-1}$), and the velocity dispersion to
be reduced to 560$\pm$68~km~s$^{-1}$ (a reduction of 0.16~dex in log
$\sigma$). The dynamical R$_{200}$ is therefore reduced from 16\%
greater than the X-ray derived value to 15\% lower -- again
inconclusive. However, in this case, the X-ray profile also exhibits
signs of disturbance in the sense that it is extended along the same
axis as the kinematic substructure (Fig. \ref{xray_overlay}). On
balance the data therefore suggest that this group is subject to an
going interaction/merger.

Evidence for merging is also present in the X-ray spectral
analysis. J1416 is the hottest and most luminous fossil group known,
with gas temperatures reaching 4 keV! It has, at larger scales, a
temperature decline seen with XMM-Newton. As shown by Khosroshahi et
al. (2006), this unusual fossil group has a
temperature ``spike'' $\sim$200~kpc from the center, followed by a
strong temperature decline at r$>$200~kpc. This spike could be due to
azimuthal temperature substructures in the inter-galactic medium. The
cooling time of 5~Gyr measured for this system is significantly below
the Hubble time for regions with the central 150~kpc (Khosroshahi
et al. 2006), but the expected level of cooling is not observed,
implying that some extra source of gas heating is in effect, maybe
shock heating due to merging.

\section{data}
In this section we present the recession velocity and apparent
magnitudes for systems with new data reported in this paper (SDSS
J0906+0301, SDSS J1045+0420, SDSS J1136+0713, SDSS J0856+0553, SDSS
J1017+0156, RX J1256.0+2556 and RX J1331.5+1108). For the
corresponding data for groups taken from the literature ( RX
J1340.5+4017 RX J1416.4+2315 and RX J1552.2+2013) see Mendes de
Oliveira \rm{et al.} (2009), Cypriano \rm{et al.} (2006) and Mendes de
Oliveira \rm{et al.} (2006), respectively. In each group the first
galaxy (with zero recession velocity ) is the BCG and the source of
the spectroscopic data used to derive recession velocities is
identified (i.e. Magellan, Gemini or the SDSS database).

\begin{table}
\caption{SDSS J0906+0301}
\begin{tabular}{|l|c|c|c|c|c|}  
   RA      &     DEC  &    RV &    g   &    r   &    i   \\
\hline   
Magellan data&          &       &        &        &        \\
09:06:38.27 &  03:01:39.1  &       0  &  16.67  &   15.54  &   15.09\\     
09:06:54.70 &  03:02:29.7  &     545  &  18.22  &   17.13  &   16.66\\    
09:06:50.68 &  03:00:02.4  &      -7  &  18.34  &   17.29  &   16.86\\      
09:06:33.72 &  03:02:00.9  &      74  &  19.73  &   18.63  &   18.18\\     
09:06:30.98 &  03:01:35.3  &    -275  &  19.74  &   18.74  &   18.29\\     
09:06:39.36 &  03:00:44.7  &    -246  &  19.78  &   18.76  &   18.28\\     
09:06:46.50 &  02:59:21.8  &     940  &  19.52  &   18.87  &   18.52\\     
09:06:43.25 &  03:02:06.0  &    -211  &  20.19  &   19.18  &   18.70\\     
09:06:46.05 &  03:03:17.6  &     167  &  20.23  &   19.25  &   18.75\\     
09:06:34.81 &  02:59:45.2  &    1120  &  20.23  &   19.43  &   18.95\\     
09:06:34.13 &  03:03:33.8  &    -224  &  20.71  &   19.75  &   19.42\\     
09:06:39.59 &  03:01:34.2  &     213  &  20.00  &   19.81  &   20.56\\     
09:06:29.70 &  03:01:03.4  &    -909  &  20.93  &   19.89  &   19.56\\     
09:06:48.28 &  03:06:09.4  &     342  &  19.21  &   18.11  &   17.67\\     
09:06:50.07 &  03:03:41.5  &    -297  &  19.41  &   18.35  &   17.94\\     
09:07:28.30 &  02:56:40.0  &     -28  &  18.97  &   18.39  &   18.05\\     
09:06:19.07 &  02:59:35.5  &    1186  &  19.92  &   18.81  &   18.27\\     
09:06:59.84 &  03:07:05.6  &     385  &  19.43  &   18.86  &   18.45\\     
09:06:51.15 &  03:00:04.7  &     362  &  19.52  &   19.05  &   18.83\\     
09:07:05.23 &  03:03:18.5  &     126  &  20.25  &   19.43  &   19.02\\     
09:06:55.76 &  03:03:15.7  &     184  &  20.08  &   19.52  &   19.14\\     
09:06:52.43 &  02:53:09.1  &      48  &  20.65  &   19.71  &   19.25\\     
09:06:17.87 &  02:58:02.5  &     437  &  20.38  &   19.76  &   19.44\\     
09:06:45.28 &  03:00:37.8  &    -174  &  21.05  &   20.02  &   19.62\\     
09:06:35.94 &  03:02:03.7  &     233  &  21.42  &   20.37  &   20.05\\     
09:06:34.54 &  03:01:47.9  &    -687  &  21.02  &   20.48  &   20.08\\     
09:06:46.05 &  03:00:44.1  &     -12  &  21.09  &   20.59  &   20.30\\     
09:06:53.43 &  02:54:17.1  &      -7  &  20.64  &   20.09  &   19.80\\     
09:06:54.78 &  03:03:00.6  &     694  &  21.18  &   20.21  &   19.77\\     
09:07:07.38 &  02:52:46.0  &      83  &  20.97  &   20.40  &   20.12\\     
09:07:22.04 &  03:03:23.2  &    1968  &  20.94  &   20.44  &   20.09\\     
09:06:54.44 &  02:57:29.8  &     392  &  21.85  &   20.86  &   20.39\\     
SDSS data   &              &          &         &          &        \\
09:07:24.35 &  02:48:25.5  &    -199  &  18.22  &   17.15  &   16.72\\     
09:07:38.51 &  02:50:20.2  &    -282  &  18.10  &   17.04  &   16.57\\     
09:07:10.05 &  02:52:34.8  &     155  &  18.39  &   17.63  &   17.18\\     
09:06:55.94 &  02:52:53.5  &     -27  &  18.53  &   17.42  &   16.94\\     
09:07:40.86 &  03:17:35.3  &      63  &  ....   &   17.53  &   17.10\\     
\hline
\end{tabular}
\label{0906}
\end{table}

\begin{table}
\caption{SDSS J1045+0420}
\begin{tabular}{|l|c|c|c|c|c|}  
   RA      &     DEC  &    RV &    g   &    r   &    i   \\
\hline   
Magellan data&          &       &        &        &        \\
10:45:48.50 &   04:20:32.5  &       0  &  17.04  &  15.83  &  15.36\\    
10:45:47.91 &   04:20:58.3  &     838  &  20.34  &  19.28  &  18.86\\    
10:45:48.48 &   04:21:15.2  &    -698  &  21.94  &  20.84  &  20.55\\    
10:45:47.49 &   04:19:40.5  &     291  &  19.99  &  18.88  &  18.40\\    
10:45:53.28 &   04:19:51.4  &    -225  &  21.52  &  20.42  &  20.10\\    
10:45:43.40 &   04:21:19.5  &     544  &  19.57  &  18.47  &  18.02\\    
10:45:43.19 &   04:19:27.7  &    1556  &  20.90  &  19.81  &  19.37\\    
10:45:53.58 &   04:19:22.6  &     689  &  19.92  &  18.83  &  18.38\\    
10:45:50.49 &   04:18:45.9  &     387  &  20.49  &  19.48  &  19.02\\    
10:45:44.37 &   04:22:05.7  &     574  &  18.95  &  17.84  &  17.38\\    
10:45:41.83 &   04:19:41.1  &     368  &  21.55  &  20.59  &  20.16\\    
10:45:43.87 &   04:22:18.5  &    -810  &  21.53  &  20.63  &  20.19\\    
10:45:56.36 &   04:21:25.7  &     489  &  19.35  &  18.28  &  17.85\\    
10:45:41.52 &   04:19:13.8  &     317  &  19.73  &  18.62  &  18.19\\    
10:45:57.30 &   04:21:03.8  &    1171  &  21.22  &  20.22  &  19.67\\    
10:45:56.48 &   04:19:18.3  &     927  &  19.83  &  19.07  &  18.63\\    
10:45:47.58 &   04:18:09.1  &     696  &  21.22  &  20.17  &  19.73\\    
10:45:43.38 &   04:18:28.3  &    -806  &  20.52  &  19.50  &  19.03\\    
10:45:58.26 &   04:21:25.4  &   -1144  &  21.09  &  20.33  &  19.94\\    
10:45:58.52 &   04:19:35.4  &     381  &  19.13  &  18.37  &  17.91\\    
10:45:54.20 &   04:23:01.9  &    -587  &  20.94  &  19.90  &  19.53\\    
10:45:55.67 &   04:22:48.0  &    -325  &  21.75  &  20.77  &  20.37\\    
10:45:39.83 &   04:22:27.7  &    -755  &  20.77  &  20.14  &  19.74\\    
10:45:59.83 &   04:19:36.8  &      -4  &  21.26  &  20.37  &  19.96\\    
10:45:42.75 &   04:17:53.8  &   -1036  &  20.95  &  20.19  &  19.81\\    
10:45:35.94 &   04:20:41.2  &    -169  &  20.54  &  19.93  &  19.69\\    
10:45:45.87 &   04:24:04.7  &    -573  &  21.46  &  20.46  &  20.07\\    
10:45:45.92 &   04:16:35.2  &    -719  &  18.97  &  18.27  &  17.89\\    
10:45:36.41 &   04:17:41.0  &     359  &  18.21  &  17.03  &  16.57\\   
10:45:58.63 &   04:24:07.3  &     798  &  19.44  &  18.36  &  17.91\\    
10:45:30.89 &   04:20:49.7  &     257  &  18.74  &  17.62  &  17.17\\     
10:45:30.25 &   04:18:41.3  &    -543  &  18.55  &  17.78  &  17.38\\  
10:46:07.15 &   04:23:33.5  &    1034  &  19.62  &  18.46  &  17.98\\    
10:45:41.60 &   04:14:44.5  &    -511  &  20.82  &  20.41  &  20.22\\    
10:45:29.90 &   04:15:21.4  &     222  &  21.77  &  20.63  &  20.25\\    
10:46:08.90 &   04:15:41.2  &    -220  &  19.84  &  18.70  &  18.18\\    
10:45:26.25 &   04:24:57.1  &     -47  &  19.23  &  18.14  &  17.65\\    
10:46:15.95 &   04:18:18.4  &    -100  &  19.98  &  19.13  &  18.72\\    
10:46:16.13 &   04:25:20.3  &    1140  &  21.02  &  20.02  &  19.44\\    
10:46:01.82 &   04:12:46.3  &    -132  &  21.14  &  20.10  &  19.61\\    
10:45:41.36 &   04:29:03.4  &     132  &  19.97  &  18.88  &  18.44\\    
10:45:27.41 &   04:27:29.2  &    -417  &  19.96  &  19.13  &  18.75\\    
10:45:46.86 &   04:11:21.2  &    -367  &  20.93  &  20.26  &  19.96\\    
10:46:15.21 &   04:27:03.3  &     791  &  20.57  &  19.93  &  19.67\\    
10:45:35.79 &   04:29:27.7  &     301  &  19.62  &  18.78  &  18.34\\    
10:46:26.09 &   04:22:21.5  &     224  &  20.61  &  19.57  &  19.09\\    
10:46:16.81 &   04:13:36.1  &    -880  &  20.83  &  20.06  &  19.55\\    
10:45:53.73 &   04:10:40.9  &     -31  &  19.55  &  18.44  &  17.97\\    
10:46:19.52 &   04:27:36.5  &    -151  &  20.17  &  19.38  &  18.93\\    
SDSS data   &    	    &          &    	 &  	   & 	   \\  
10:46:49.97 &   04:19:38.1  &    1875  &  18.63  &  17.71  &  17.23\\    
10:46:35.28 &   04:02:56.3  &     -13  &  18.72  &  17.66  &  17.19\\    
10:45:02.05 &   04:01:59.9  &    -597  &  18.75  &  17.59  &  17.14\\    
10:47:07.62 &   04:30:29.9  &     -18  &  18.85  &  17.73  &  17.22\\    
10:47:22.44 &   04:16:30.1  &    -102  &  18.27  &  17.09  &  16.48\\    
10:44:46.30 &   03:53:06.9  &    -622  &  17.58  &  16.42  &  15.94\\    
10:44:53.91 &   03:51:28.5  &    -993  &  18.59  &  17.56  &  17.11\\    
10:43:50.96 &   04:36:15.3  &    -855  &  17.85  &  16.68  &  16.20\\    
10:44:00.45 &   04:42:41.9  &    -686  &  18.85  &  17.68  &  17.16\\    
10:46:45.93 &   04:52:46.8  &     308  &  18.73  &  17.70  &  17.30\\    
\hline
\end{tabular}
\label{1045}
\end{table}

\begin{table}
\caption{SDSS J1136+0713}
\begin{tabular}{|l|c|c|c|c|c|}  
   RA      &     DEC  &    RV &    g   &    r   &    i   \\
\hline   
Magellan data&          &       &        &        &        \\
11:36:23.71 &   07:13:37.5   &      0  &  15.90 &   14.92 &   14.46\\
11:36:25.87 &   07:13:19.5   &    187  &  19.45 &   18.56 &   18.10\\
11:36:21.15 &   07:14:03.7   &    565  &  20.88 &   20.02 &   19.59\\
11:36:26.82 &   07:13:40.7   &    488  &  19.68 &   18.75 &   18.34\\
11:36:27.02 &   07:13:30.6   &   -239  &  19.30 &   18.31 &   17.92\\
11:36:22.45 &   07:12:49.9   &    698  &  19.04 &   18.00 &   17.53\\
11:36:26.78 &   07:14:25.3   &    102  &  20.22 &   19.25 &   18.86\\
11:36:18.78 &   07:13:53.8   &   -637  &  19.06 &   18.07 &   17.60\\
11:36:18.67 &   07:14:00.0   &    150  &  18.85 &   17.84 &   17.38\\
11:36:29.45 &   07:14:22.3   &  -1030  &  18.40 &   17.69 &   17.30\\
11:36:29.37 &   07:14:26.5   &    117  &  19.62 &   18.66 &   18.16\\
11:36:16.92 &   07:12:51.2   &   -136  &  20.23 &   19.33 &   18.91\\
11:36:30.28 &   07:14:30.8   &   -844  &  18.31 &   17.35 &   16.93\\
11:36:13.44 &   07:13:18.6   &   -228  &  20.72 &   19.82 &   19.40\\
11:36:34.05 &   07:13:08.6   &   -162  &  19.15 &   18.16 &   17.71\\
11:36:13.93 &   07:12:34.9   &   1060  &  19.19 &   18.24 &   18.00\\
11:36:13.52 &   07:14:25.2   &    469  &  20.75 &   20.06 &   19.67\\
11:36:34.50 &   07:13:08.7   &   -387  &  20.57 &   19.60 &   19.18\\
11:36:30.19 &   07:15:50.2   &   -225  &  19.09 &   18.61 &   18.30\\
11:36:19.95 &   07:16:11.7   &   -283  &  18.97 &   18.03 &   17.63\\
11:36:12.91 &   07:12:55.5   &   -302  &  18.15 &   17.17 &   16.75\\ 
11:36:31.90 &   07:15:31.3   &   -805  &  20.99 &   20.37 &   20.05\\
11:36:33.70 &   07:11:55.8   &   -341  &  19.34 &   19.04 &   18.73\\
11:36:12.94 &   07:15:23.5   &    152  &  20.26 &   19.28 &   18.90\\
11:36:33.63 &   07:11:32.5   &    308  &  18.45 &   17.49 &   17.07\\
11:36:13.87 &   07:11:23.3   &   -425  &  20.38 &   19.43 &   19.01\\
11:36:37.31 &   07:12:29.6   &   -213  &  18.38 &   17.51 &   17.09\\
11:36:08.36 &   07:13:45.5   &   -169  &  16.50 &   15.49 &   15.02\\
11:36:08.50 &   07:12:42.8   &    186  &  19.34 &   18.38 &   17.95\\
11:36:14.45 &   07:10:28.6   &    712  &  18.97 &   18.01 &   17.54\\
11:36:09.64 &   07:11:29.2   &    929  &  20.63 &   19.88 &   19.40\\
11:36:37.14 &   07:16:05.8   &    766  &  21.69 &   20.83 &   20.75\\
11:36:06.84 &   07:13:52.4   &   -243  &  16.97 &   15.95 &   15.50\\
11:36:39.35 &   07:11:38.9   &    410  &  19.37 &   18.42 &   18.00\\
11:36:05.22 &   07:11:57.8   &    134  &  17.59 &   16.60 &   16.15\\
11:36:41.66 &   07:16:41.6   &   -148  &  20.19 &   19.19 &   18.70\\
11:36:48.67 &   07:13:09.2   &   -983  &  19.51 &   18.62 &   18.15\\
11:36:42.03 &   07:18:13.9   &    -26  &  17.87 &   16.87 &   16.42\\
11:35:57.72 &   07:14:19.4   &   -708  &  21.12 &   20.41 &   20.16\\
11:36:50.22 &   07:15:45.7   &    234  &  20.78 &   19.87 &   19.51\\
11:36:31.27 &   07:20:38.1   &    190  &  19.72 &   18.83 &   18.27\\
11:35:59.90 &   07:08:35.5   &    480  &  18.82 &   18.36 &   18.04\\
11:36:48.73 &   07:19:39.8   &    183  &  18.95 &   18.12 &   17.67\\
11:37:01.13 &   07:12:29.1   &   -417  &  19.07 &   18.45 &   18.12\\
11:35:45.01 &   07:15:41.4   &    -64  &  21.36 &   20.56 &   20.23\\
11:36:46.73 &   07:23:00.2   &    -28  &  19.60 &   18.82 &   18.36\\
11:35:36.83 &   07:09:06.3   &    805  &  18.28 &   17.35 &   16.94\\
11:35:34.30 &   07:10:53.4   &    852  &  19.31 &   18.32 &   17.88\\  
\hline      
\end{tabular}
\label{1136}
\end{table}

\begin{table}
\caption{SDSS J0856+0553}
\begin{tabular}{|l|c|c|c|c|c|}  
   RA      &     DEC  &    RV &    g   &    r   &    i   \\
\hline   
Magellan data&          &       &        &        &        \\
08:56:40.72 &   05:53:47.3   &      0  &  16.05 &   15.02  &  14.57\\
08:56:40.71 &   05:53:04.0   &    651  &  19.83 &   18.90  &  18.44\\
08:56:43.68 &   05:53:33.8   &   -104  &  19.43 &   18.35  &  17.99\\
08:56:38.61 &   05:52:52.9   &     11  &  19.56 &   18.62  &  18.19\\
08:56:40.09 &   05:54:52.5   &     67  &  19.40 &   18.42  &  18.00\\
08:56:38.75 &   05:52:28.5   &   -634  &  20.72 &   19.80  &  19.35\\
08:56:34.91 &   05:54:17.7   &    -91  &  20.13 &   19.26  &  18.83\\
08:56:40.57 &   05:55:20.5   &  -1226  &  18.89 &   18.31  &  17.91\\
08:56:47.32 &   05:53:59.5   &    213  &  20.15 &   19.24  &  18.81\\
08:56:41.34 &   05:55:34.4   &     80  &  18.88 &   17.88  &  17.44\\
08:56:37.31 &   05:51:59.2   &   -803  &  18.60 &   17.64  &  17.18\\
08:56:38.04 &   05:51:49.3   &   -182  &  19.78 &   18.85  &  18.41\\
08:56:34.22 &   05:55:40.4   &   -420  &  19.58 &   18.81  &  18.40\\
08:56:45.66 &   05:51:36.2   &   -393  &  20.66 &   19.87  &  19.45\\
08:56:44.69 &   05:51:28.0   &   -646  &  19.35 &   18.41  &  17.98\\
08:56:35.97 &   05:51:32.2   &   -588  &  18.87 &   17.93  &  17.47\\
08:56:29.10 &   05:54:39.6   &   -554  &  20.14 &   19.30  &  18.93\\
08:56:28.55 &   05:53:55.1   &    141  &  21.56 &   20.67  &  20.37\\
08:56:47.62 &   05:56:30.8   &   1119  &  20.85 &   19.97  &  19.56\\
08:56:52.51 &   05:55:10.4   &    107  &  19.58 &   18.62  &  18.19\\
08:56:29.49 &   05:51:40.9   &    420  &  19.60 &   18.72  &  18.30\\
08:56:39.61 &   05:50:11.6   &    154  &  19.65 &   18.74  &  18.30\\
08:56:30.27 &   05:51:07.8   &   -418  &  21.13 &   20.34  &  19.86\\
08:56:51.01 &   05:56:34.2   &   -175  &  19.55 &   18.60  &  18.22\\
08:56:31.75 &   05:50:43.6   &    208  &  20.62 &   19.59  &  19.22\\
08:56:23.73 &   05:53:19.9   &    275  &  18.86 &   17.83  &  17.33\\
08:56:25.67 &   05:56:04.2   &   -390  &  21.24 &   20.41  &  20.01\\
08:56:24.24 &   05:49:37.9   &     39  &  20.54 &   19.65  &  19.28\\
08:57:02.52 &   05:56:26.5   &    522  &  19.75 &   19.49  &  19.27\\
08:56:35.46 &   05:47:18.1   &   -177  &  19.69 &   19.24  &  18.94\\
08:57:03.70 &   05:57:24.2   &   -166  &  19.53 &   18.56  &  18.03\\
08:56:11.67 &   05:52:55.9   &   -625  &  21.38 &   20.45  &  20.13\\
08:56:59.40 &   05:48:07.3   &    730  &  18.98 &   18.04  &  17.59\\
08:56:50.78 &   05:46:47.5   &    -67  &  19.72 &   18.78  &  18.37\\
08:56:56.09 &   05:47:22.9   &   -594  &  19.67 &   18.91  &  18.51\\
08:56:29.71 &   05:46:22.4   &   -178  &  21.59 &   20.63  &  20.14\\
08:57:12.61 &   05:51:31.1   &    171  &  20.81 &   20.16  &  19.77\\
08:57:13.75 &   05:52:34.4   &    187  &  19.70 &   18.98  &  18.51\\
08:56:22.76 &   05:46:30.2   &    449  &  19.62 &   18.73  &  18.30\\
08:56:32.03 &   05:45:00.2   &    820  &  20.47 &   19.71  &  19.43\\
08:56:13.22 &   05:59:44.9   &   -335  &  20.56 &   20.12  &  19.86\\
08:57:08.63 &   05:59:50.9   &    184  &  20.13 &   19.15  &  18.80\\
08:57:17.72 &   05:54:32.4   &   -443  &  19.33 &   18.52  &  18.04\\
08:56:02.99 &   05:54:52.1   &    243  &  21.64 &   20.70  &  20.47\\
08:56:39.86 &   06:04:05.9   &   -273  &  20.97 &   20.37  &  20.07\\
08:57:04.71 &   05:44:16.1   &   -515  &  20.50 &   19.89  &  19.63\\
08:55:53.45 &   05:54:04.5   &    220  &  20.84 &   19.75  &  19.32\\
08:56:49.02 &   05:40:39.5   &   1068  &  21.93 &   20.94  &  20.86\\
SDSS data   &		     &         &       	&          &       \\
08:56:54.63 &   05:49:35.6   &    291  &  18.42 &   17.54  &  17.11\\
08:56:55.79 &   05:56:29.8   &    926  &  18.03 &   17.14  &  16.73\\
08:56:30.82 &   05:57:20.6   &    329  &  17.96 &   17.06  &  16.65\\
08:56:41.49 &   05:51:37.8   &   1394  &  18.31 &   17.27  &  16.81\\
08:56:39.80 &   05:55:26.2   &    877  &  18.51 &   17.53  &  17.09\\
08:56:50.01 &   05:48:38.7   &    152  &  17.73 &   16.75  &  16.32\\
08:56:25.57 &   05:58:04.6   &   -275  &  18.50 &   17.52  &  17.09\\
08:56:23.21 &   05:49:49.4   &    307  &  17.74 &   16.69  &  16.22\\
08:57:06.03 &   05:51:54.4   &   -190  &  18.19 &   17.24  &  16.80\\
08:56:07.76 &   05:54:45.0   &    264  &  17.22 &   16.27  &  15.85\\
08:56:33.00 &   06:01:50.3   &   -199  &  17.63 &   16.65  &  16.21\\
08:57:09.34 &   05:48:26.5   &    307  &  17.74 &   16.73  &  16.27\\
08:56:07.24 &   05:55:53.2   &     87  &  17.92 &   16.94  &  16.52\\
08:57:25.80 &   05:57:57.8   &    397  &  17.29 &   16.35  &  15.93\\
\hline
\end{tabular}
\label{0856a}
\end{table}

\begin{table}
\caption{SDSS J0856+0553 cont.}
\begin{tabular}{|l|c|c|c|c|c|}  
   RA      &     DEC  &    RV &    g   &    r   &    i   \\
\hline   
SDSS data   &                &         &        &          &       \\

08:56:49.61 &   05:41:33.5   &    128  &  18.59 &   17.70  &  17.29\\
08:55:57.99 &   05:52:14.8   &    -24  &  17.51 &   16.56  &  16.11\\
08:55:42.60 &   05:55:42.5   &    -84  &  18.26 &   17.31  &  16.85\\
08:55:50.98 &   06:03:31.5   &   -147  &  17.21 &   16.55  &  16.15\\
08:57:25.69 &   05:41:41.1   &     35  &  17.54 &   16.52  &  16.07\\
08:55:05.94 &   05:58:06.8   &   1287  &  17.99 &   17.21  &  16.77\\
08:56:42.63 &   05:28:19.4   &    109  &  18.71 &   17.67  &  17.17\\
08:58:17.91 &   06:05:34.2   &    221  &  18.31 &   17.73  &  17.40\\
08:58:30.66 &   05:49:31.8   &   -111  &  17.99 &   16.98  &  16.50\\
08:55:08.10 &   06:09:53.8   &    -76  &  17.66 &   17.07  &  16.54\\
08:58:33.14 &   05:48:08.1   &    199  &  18.04 &   16.95  &  16.47\\
08:58:45.57 &   05:51:03.5   &    183  &  17.31 &   16.46  &  15.98\\
08:58:03.69 &   06:19:04.6   &    384  &  18.36 &   17.75  &  17.49\\
08:58:08.74 &   06:18:16.4   &   -146  &  17.83 &   16.80  &  16.31\\
08:58:05.20 &   06:20:12.7   &    -65  &  17.56 &   16.82  &  16.39\\
08:54:35.32 &   05:39:41.9   &     28  &  17.19 &   16.26  &  15.82\\
08:54:29.11 &   05:43:39.2   &   1411  &  18.46 &   17.55  &  17.17\\
08:57:29.42 &   06:26:56.2   &   -144  &  18.48 &   17.71  &  17.32\\
08:55:17.80 &   05:24:53.4   &    720  &  17.71 &   16.94  &  16.53\\
08:56:34.13 &   06:29:39.7   &    240  &  18.03 &   17.20  &  16.67\\
08:56:12.98 &   06:33:29.7   &   1127  &  18.64 &   18.09  &  17.66\\
08:56:48.24 &   06:34:32.4   &    856  &  18.67 &   17.82  &  17.30\\
08:55:27.47 &   06:30:23.8   &   1409  &  16.84 &   16.22  &  15.84\\
08:56:38.82 &   05:53:33.1   &   -150  &  18.85 &   17.83  &  17.35\\
\hline
\end{tabular}
\label{0856b}
\end{table}

\begin{table}
\caption{SDSS 1017+0156}
\begin{tabular}{|l|c|c|c|c|c|}  
   RA      &     DEC  &    RV &    g   &    r   &    i   \\
\hline   
Magellan data&          &       &        &        &        \\
10:17:45.57 &  01:56:45.8   &      0 &   16.36 &    15.28  &   14.78\\
10:17:42.06 &  01:53:47.8   &   -747 &   18.18 &    17.16  &   16.74\\
10:17:55.52 &  01:54:34.8   &    445 &   19.06 &    17.97  &   17.50\\
10:17:38.88 &  01:53:28.9   &    327 &   19.12 &    18.05  &   17.61\\
10:17:43.48 &  01:57:56.0   &  -1298 &   19.11 &    18.07  &   17.64\\
10:17:55.90 &  01:55:16.4   &     80 &   19.36 &    18.59  &   18.24\\
10:17:47.14 &  01:54:44.6   &    279 &   19.70 &    18.67  &   18.23\\
10:17:32.07 &  01:55:29.6   &   -459 &   21.00 &    19.92  &   19.43\\
10:18:00.06 &  01:56:34.6   &   -193 &   20.98 &    20.00  &   19.64\\
10:17:39.44 &  01:53:44.2   &   -759 &   21.49 &    20.99  &   20.69\\
10:18:36.01 &  02:01:58.1   &   1179 &   19.08 &    18.07  &   17.55\\
10:18:14.74 &  01:59:15.8   &    295 &   19.00 &    18.43  &   18.01\\
10:18:16.78 &  02:06:01.8   &   1365 &   19.40 &    18.62  &   18.19\\
10:17:46.32 &  02:02:13.8   &    157 &   20.08 &    19.30  &   18.80\\
10:18:00.24 &  02:00:51.3   &    350 &   20.27 &    19.61  &   19.23\\
10:18:33.57 &  01:55:12.5   &    253 &   20.32 &    19.63  &   19.25\\
10:18:32.17 &  01:59:50.0   &   1508 &   20.81 &    20.11  &   19.73\\
10:17:28.87 &  01:55:16.3   &   -658 &   21.26 &    20.54  &   20.19\\
10:17:24.90 &  01:45:54.4   &    982 &   21.82 &    20.72  &   20.23\\
10:17:51.72 &  01:57:27.6   &   -513 &   19.15 &    17.99  &   17.45\\
10:17:45.08 &  01:54:37.0   &   -198 &   19.26 &    18.21  &   17.80\\
10:17:42.79 &  01:57:08.1   &    307 &   19.41 &    18.32  &   17.87\\
10:17:55.73 &  01:59:27.6   &   -300 &   20.63 &    19.60  &   19.12\\
10:17:57.54 &  01:55:13.0   &    562 &   21.50 &    20.57  &   19.99\\
10:18:12.59 &  01:51:42.0   &    137 &   20.31 &    19.26  &   18.76\\
10:18:12.71 &  01:52:57.3   &     93 &   20.23 &    19.66  &   19.41\\
10:17:04.75 &  01:52:47.4   &   1177 &   20.57 &    19.89  &   19.60\\
10:18:17.90 &  01:57:32.7   &    309 &   20.54 &    19.91  &   19.44\\
10:17:32.26 &  01:52:26.3   &     97 &   21.78 &    20.74  &   20.45\\
10:16:51.08 &  01:56:04.0   &    257 &    .... &     ....  &    ....\\
SDSS data   &               &        &         &           &        \\
10:17:16.15 &  02:16:22.6   &   1117 &   17.41 &    16.38  &   15.91\\
10:17:43.32 &  01:38:06.4   &   1185 &   17.97 &    17.68  &   17.39\\
10:17:45.73 &  02:14:06.7   &    936 &   18.44 &    17.40  &   16.97\\
10:18:37.23 &  02:03:42.9   &   1221 &   17.94 &    16.83  &   16.31\\
10:18:50.50 &  02:04:22.8   &   1124 &   18.51 &    17.42  &   16.93\\
10:19:01.74 &  01:56:19.9   &    -87 &   18.37 &    17.55  &   17.06\\
\hline
\end{tabular}
\label{1017}
\end{table} 

\begin{table}
\caption{RX J1256.0+2556}
\begin{tabular}{|l|c|c|c|c|c|}  
   RA      &     DEC  &    RV &    g   &    r   &    i   \\
\hline   
Gemini data&          &       &        &        &        \\
12:56:02.27 & 25:56:37.2  &      0 &   18.22 &    16.76  &     16.29\\       
12:55:54.88 & 25:57:59.0  &  -1228 &   22.09 &     ....  &     20.99\\         
12:55:55.59 & 25:56:33.8  &  -1111 &   21.00 &     ....  &     19.40\\         
12:56:01.97 & 25:56:40.4  &  -1029 &   20.33 &     ....  &     18.75\\         
12:55:55.55 & 25:57:10.8  &   -876 &   21.18 &     ....  &     19.56\\         
12:55:51.19 & 25:57:33.5  &   -817 &   22.74 &     ....  &     21.35\\         
12:56:02.55 & 25:56:50.0  &   -790 &   20.68 &     ....  &     19.37\\         
12:55:54.98 & 25:58:24.8  &   -612 &   20.13 &     ....  &     18.60\\         
12:56:09.04 & 25:55:47.5  &   -479 &   20.62 &     ....  &     19.57\\         
12:56:08.88 & 25:55:57.9  &   -411 &   20.38 &     ....  &     18.98\\         
12:55:54.33 & 25:56:14.5  &   -187 &   20.71 &     ....  &     19.53\\         
12:55:57.90 & 25:58:19.5  &   -148 &   19.25 &     ....  &     17.63\\         
12:56:01.09 & 25:57:02.2  &   -121 &   20.61 &     ....  &     18.94\\         
12:56:03.29 & 25:54:37.1  &     75 &   19.84 &     ....  &     18.10\\         
12:56:00.57 & 25:57:22.8  &    138 &   21.61 &     ....  &     20.06\\         
12:56:02.34 & 25:55:14.2  &    187 &   20.50 &     ....  &     18.91\\         
12:56:01.73 & 25:55:18.3  &    229 &   19.91 &     ....  &     18.45\\         
12:56:01.70 & 25:56:28.9  &    259 &   20.71 &     ....  &     19.41\\         
12:55:54.43 & 25:58:10.4  &    276 &   21.94 &     ....  &     20.38\\         
12:56:02.67 & 25:56:32.8  &    479 &   20.25 &     ....  &     18.68\\         
12:56:09.34 & 25:56:09.6  &    520 &   22.06 &     ....  &     20.35\\         
12:56:03.17 & 25:56:24.5  &    749 &   22.47 &     ....  &     19.98\\         
12:56:00.86 & 25:55:21.4  &    775 &   21.28 &     ....  &     19.90\\         
SDSS data   &             &        &         &           &          \\
12:55:04.55 & 26:02:54.7  &   -982 &   18.87 &    17.50  &     16.97\\         
12:54:53.39 & 25:50:56.7  &    -18 &   18.52 &    17.12  &     16.59\\         
12:54:38.82 & 25:52:13.5  &   -825 &   18.90 &    17.46  &     16.96\\         
12:56:07.18 & 25:57:23.9  &    759 &   20.33 &     ....  &     18.65\\         
12:56:03.48 & 25:57:02.2  &   -921 &   20.96 &     ....  &     19.20\\         
12:55:53.01 & 25:57:23.9  &   -872 &   19.93 &     ....  &     18.04\\         
12:55:56.30 & 25:55:56.3  &    152 &    .... &     ....  &      ....\\         
12:56:02.24 & 25:56:32.0  &    565 &   19.68 &     ....  &     17.82\\         
\hline
\end{tabular}
\label{1256}
\end{table}

\begin{table}
\caption{RX J1331.5+1108}
\begin{tabular}{|l|c|c|c|c|c|}  
   RA      &     DEC  &    RV &    g   &    r   &    i   \\
\hline   
Gemini data&          &       &        &        &        \\
13:31:29.7 &  11:07:6 &     0 &  16.36 &  15.40 &  14.95 \\
13:31:33.4 &  11:05:5 &    51 &  25.12 &  22.16 &  22.07 \\
13:31:22.5 &  11:07:2 &   359 &  19.88 &  19.57 &  19.33 \\
13:31:34.6 &  11:08:0 &   666 &  23.14 &  21.31 &  20.53 \\
13:31:24.0 &  11:08:2 &  -309 &  18.71 &  17.84 &  17.46 \\
SDSS data  &          &       &        &        &        \\
13:31:36.4 &  11:13:0 &   495 &  17.07 &  16.47 &  16.11 \\
13:31:41.5 &  11:06:5 &  -328 &  18.03 &  17.25 &  16.84 \\
13:31:14.3 &  11:08:2 &   -88 &  17.79 &  16.92 &  16.45 \\
13:31:09.9 &  11:05:2 &   163 &  17.15 &  16.28 &  15.91 \\
13:31:26.0 &  11:05:3 &   412 &  18.25 &  17.55 &  17.10 \\
\hline
\end{tabular}
\label{1331}
\end{table}

\end{appendix}

\label{lastpage}
\end{document}